\documentclass[
preprint,
nofootinbib,
 amsmath,amssymb,
 aps,
]{revtex4-2}

\usepackage{graphicx}
\usepackage{dcolumn}
\usepackage{bm}
\usepackage{amsmath,amssymb}
\usepackage{extpfeil} 
\usepackage{mathtools}
\usepackage{ulem}
\usepackage{stmaryrd}
\usepackage{hyperref}
\usepackage{xcolor}
\hypersetup{
    colorlinks=true,
    linkcolor=blue,  
    urlcolor=cyan     
}

\newcommand{\defeq}{:=}
\newcommand{\eqdef}{=:}
\newcommand{\Vc}[1]{\boldsymbol{#1}}
\newcommand{\dd}{\mathrm{d}}
\newcommand{\dt}{\dd t}

\newcommand{\Transpose}{\top}
\newcommand{\IncMatrix}{\mathbb{B}}

\newcommand{\Real}{\mathbb{R}}
\newcommand{\Integer}{\mathbb{Z}}
\newcommand{\kcoef}{k}
\newcommand{\kcoefop}{\mathfrak{k}}

\newcommand{\stoiMatrix}{\mathbb{S}}

\newcommand{\indMatrix}{\mathbb{B}}

\newcommand{\sutilF}{\mathcal{U}}
\newcommand{\futilF}{\mathcal{F}}
\newcommand{\costF}{\mathcal{C}}
\newcommand{\valueF}{V}

\newcommand{\KLD}{\mathcal{D}}
\newcommand{\EPR}{\dot{\Sigma}}

\newcommand{\eqnref}[1]{Eq. \eqref{#1}}

\newcommand{\Volume}{\Omega}

\newcommand{\setX}{\mathcal{X}}

\newcommand{\setR}{\mathcal{R}}
\newcommand{\Img}{\mathrm{Im}}
\newcommand{\Ker}{\mathrm{Ker}}

\newcommand{\flux}{j}
\newcommand{\fluxop}{\mathfrak{j}}
\newcommand{\tf}{f}
\newcommand{\tff}{\eta}
\newcommand{\gConst}{R}
\newcommand{\dissip}{\Psi}

\newcommand{\diag}{\mathrm{diag}}
\newcommand{\Hamiltonian}{\mathcal{H}}
\newcommand{\Lagrangian}{\mathcal{L}}
\newcommand{\gmean}{\mathfrak{m}}
\newcommand{\const}{\mathrm{const.}}

\usepackage{amsthm}

\begin{document}

\preprint{APS/123-QED}

\title{Optimal Control of Chemical Reaction Networks with Thermodynamic Cost}

\author{Tetsuya J. Kobayashi}
 \affiliation{Institute of Industrial Science, The University of Tokyo, Tokyo 153-8505, Japan}
 \affiliation{Universal Biology Institute, The University of Tokyo, Tokyo 113-8654, Japan.}
\email{tetsuya@mail.crmind.net}
 
\author{Alix Robin}
\thanks{These authors contributed equally to this work.}
\affiliation{Ecole Normale Superieure in ENS-PSL}

\author{Shuhei A. Horiguchi}
\thanks{These authors contributed equally to this work.}
\affiliation{Nano Life Science Institute (NanoLSI), Kanazawa University}



\date{\today}

\begin{abstract}
Thermodynamic control seeks to realize desired functions at minimal thermodynamic cost, yet existing frameworks remain largely confined to slow driving regimes or state-to-state transport. Here we establish a thermodynamic optimal-control theory for Markov jump processes (MJPs) and chemical reaction networks (CRNs) that accommodates objectives over both states and reaction fluxes beyond slow-driving limit. By deriving the thermodynamic Hamilton--Jacobi--Bellman equation, we demonstrate that the optimal dynamics take the form of a perturbed generalized-gradient flow driven by the value function. A convex representation of the entropy-production rate via generalized means identifies its Legendre--Fenchel dual as the dissipation function. The resulting dynamics shares the thermodynamic structure of CRNs and admits an information-geometric characterization. We further derive a thermodynamic Pontryagin equation by recasting the control problem in terms of thermodynamic forces, providing a complementary description of optimal trajectories and clarifying connections to thermodynamic optimal transport. 
Our results establish a thermodynamic and geometric foundation for controlling MJPs and CRNs beyond mere transport, opening avenues for optimizing biological processes and finite-time thermodynamic operations in far-from-equilibrium regimes.
\end{abstract}

\maketitle


\section{Introduction}
\subsection{Challenge in Thermodynamics}
A central challenge in thermodynamics is designing protocols that drive systems to perform desired functions at minimal thermodynamic cost. 
This challenge spans diverse applications, from engineering efficient heat engines to elucidating the design principles of biochemical reaction networks and the fundamental physical limits of computation \cite{mcgrath2026BiophysicsRev.,datta2022Phys.Rev.X,sahin2023NatCommun,wolpert2024Proc.Natl.Acad.Sci.}. 
Recent advances in stochastic thermodynamics \cite{sekimoto1998ProgTheorPhys,peliti2021,seifert2025,avanzini2024SciPostPhys.Lect.Notes} have overcome the limitations of classical finite-time thermodynamics \cite{curzon1975Am.J.Phys.,andresen2026Entropy} by providing a framework for quantifying work, heat, and entropy production under externally driven operations at both the distribution and trajectory levels. 

Building on this theoretical foundation, extensive efforts have been dedicated to characterizing thermodynamic bounds and identifying optimal protocols for canonical tasks, such as trajectory tracking and state-to-state transitions, under various constraints and assumptions \cite{deffner2020EPL,blaber2023J.Phys.Commun.,alvarado2026Annu.Rev.Condens.MatterPhys.}.

\subsection{Near-equilibrium thermodynamic control and thermodynamic geometry}
In the near-equilibrium regime where linear response theory holds, slowly driven finite-time protocols that minimize dissipation correspond to geodesics on a Riemannian manifold endowed with a generalized friction metric \cite{crooks2007Phys.Rev.Lett.a}. 
This geometric formulation casts the search for optimal, minimum-dissipation protocols into a problem governed by the thermodynamic metric \cite{sivak2012Phys.Rev.Lett.,zulkowski2012Phys.Rev.E}.

Since its inception\cite{salamon1980J.Chem.Phys.,salamon1983Phys.Rev.Lett.a,andresen1996RevueGeneraledeThermiquea} via the integration of finite-time thermodynamics\cite{ma2024,andresen2011Angew.Chem.Int.Ed.,andresen1977J.Chem.Phys.,andresen1977Phys.Rev.Ac,salamon1977Phys.Rev.Aa} and Weinhold--Ruppeiner equilibrium geometry\cite{weinhold1975J.Chem.Phys.a,ruppeiner1979Phys.Rev.A},  the geometric approach has been refined theoretically \cite{li2025Phys.Rev.E,plastino2026Mod.Phys.Lett.A,watanabe2026} and applied to a diverse range of model systems \cite{blaber2023J.Phys.Commun.,wareham2025Phys.Rev.E}. 
Moreover, leveraging the Hatano--Sasa excess heat, the scope of the framework has been expanded from a near-equilibrium tracking problem to transitions between non-equilibrium steady states \cite{zulkowski2013PLOSONE}, evolving into non-equilibrium quasi-static control \cite{kamp2026Phys.Rev.E,sawchuk2026Phys.Rev.Res.,matsuo2022Phys.Rev.E,wang2024Phys.Rev.Lett.}.

Additionally, shortcuts to adiabaticity or to isothermality have been developed, which realize a target sequence of equilibrium or steady states in finite time by introducing auxiliary controls \cite{guery-odelin2019Rev.Mod.Phys.,patra2017NewJ.Phys.,guery-odelin2023Rep.Prog.Phys.,ilker2022Phys.Rev.X}.

\subsection{Far from equilibrium transport and its extensions}
While powerful, these frameworks are restricted to near-equilibrium or steady-state regimes. 
Two pioneering studies explored possible ways to resolve this difficulty\cite{aurell2011Phys.Rev.Lett.a,muratore-ginanneschi2013J.Phys.A:Math.Theor.}; 
Notably, the work by Aurell \textit{et al.}, which showed that finding a transport protocol with minimizing excess work and heat can be mapped to an optimal transport problem \cite{aurell2011Phys.Rev.Lett.a,aurell2012JStatPhys}, established a theoretical foundation for far-from-equilibrium transport. 
After a gap of about a decade since its proposal, theoretical and experimental efforts have rapidly accumulated in recent years, particularly for overdamped Langevin systems \cite{chen2020IEEETrans.Autom.Control,nakazato2021Phys.Rev.Res.,movillamiangolarra2022IEEEControlSyst.Lett.a,chennakesavalu2023Phys.Rev.Lett.,ito2024Info.Geo.,kamijima2025Phys.Rev.Res.,klinger2025Phys.Rev.E,tojo2025,oikawa2025NatCommun}. 
These insights have subsequently been extended to discrete Markov Jump Processes (MJPs) and chemical reaction networks (CRNs)\cite{dechant2022J.Phys.A:Math.Theor.,vanvu2023Phys.Rev.Xa,nagayama2025Phys.Rev.Res.,nagayama2025Phys.Rev.Res.a,meer2026}.
Moreover, optimal transport offers a Wasserstein-geometric perspective that allows the derivation of various bounds in general thermodynamic settings.

Nevertheless, existing approaches remain limited to simple transport problems (Fig. \ref{fg:outline} (a)). In practical thermodynamic control, however, one often faces demands that extend far beyond mere transport.
Specifically, maximizing target transitions and fluxes (Fig. \ref{fg:outline} (b))\cite{brown2017Proc.Natl.Acad.Sci.,andrieux2023,yang2026NatCommun} is essential for optimizing work extraction\cite{schmiedl2007EPL}, reaction fluxes\cite{fleming2012JournalofTheoreticalBiology,gagrani2026J.Stat.Mech.}, and free energy transduction\cite{pinero2024Phys.Rev.Res.,bilancioni2025NatCommun}.
Furthermore, confining a system to a prescribed region is often crucial to the stability of thermodynamic systems (Fig. \ref{fg:outline} (c))\cite{horowitz2017Phys.Rev.E,bilancioni2025NatCommun}.
Even though such problems, especially flux optimization with minimal costs, have been addressed in the context of biochemical machines and networks operating at steady state, their integration with transport phenomena is yet to be achieved.

\subsection{Optimal Control and Thermodynamics}
Optimal control theory, which generalizes variational calculus, offers a natural mathematical setting for treating such diverse requirements without assuming slow driving \cite{bryson1975,liberzon2011,bertsekas2012,alvarado2026Annu.Rev.Condens.MatterPhys.,alvarado2026Annu.Rev.Condens.MatterPhys.}. 
Optimal control theory incorporates a variety of requirements and constraints in the form of utilities and costs into a single objective functional and provides two principled ways to solve the resulting functional optimization problems. On the one hand,  the Hamilton--Jacobi--Bellman (HJB) equation, grounded in Bellman's principle of optimality, provides the state-time global landscape of desirability that the optimal trajectory should follow.
On the other hand, the Hamiltonian dynamics derived from Pontryagin's maximum principle can directly generate the state-time local individual optimal trajectory and dynamic driving control without the knowledge of the global desirability landscape\cite{blaber2023J.Phys.Commun.}. These two together provide a coherent view of the global and local  aspects of the optimal control problem.
Yet, despite their solid and profound mathematical basis,  in the context of thermodynamic control, variational and optimal-control methods have largely been used as mere technical tools to derive problem-specific optimal protocols for individual setups\cite{schmiedl2007Phys.Rev.Lett.,muratore-ginanneschi2013J.Phys.A:Math.Theor.,bo2013EPL,zhong2022Phys.Rev.E,monter2025Proc.Natl.Acad.Sci.,kurokawa2026,meibohm2026,panizon2026},  rather than as a theoretical foundation for thermodynamic control comparable to thermodynamic geometry or optimal transport \cite{pavon1989ApplMathOptim}. 

In addition, its application has been heavily skewed toward continuous diffusion processes described by the Fokker-Planck equation, since existing optimal control theory is predominantly built on such processes. Extensions to discrete-state MJPs and CRNs, which possess mathematical structures distinct from their continuous counterpart, remain limited. Indeed, recent studies on thermodynamic control of CRNs are restricted to the slow-driving limit \cite{zhang2025Phys.Rev.Lett.,mohite2026}.
A notable exception is the work of Muratore-Ginanneschi \textit{et al.}, which treated thermodynamic transport in MJPs within an optimal-control formulation \cite{muratore-ginanneschi2013JStatPhys,remlein2021Phys.Rev.E}. However, their approach relies on the coordinate transformation from the probability variable $p$  to a wave-function-like variable $\phi=\sqrt{p}$ to employ the optimal control techniques for quantum-mechanical systems. This construction differs substantially from the standard optimal-control formulation, rendering it non-trivial to extend it to other contexts, such as CRNs. 

\subsection{Outline of this work}
\begin{figure}
    \centering
    \includegraphics[width=0.95\linewidth]{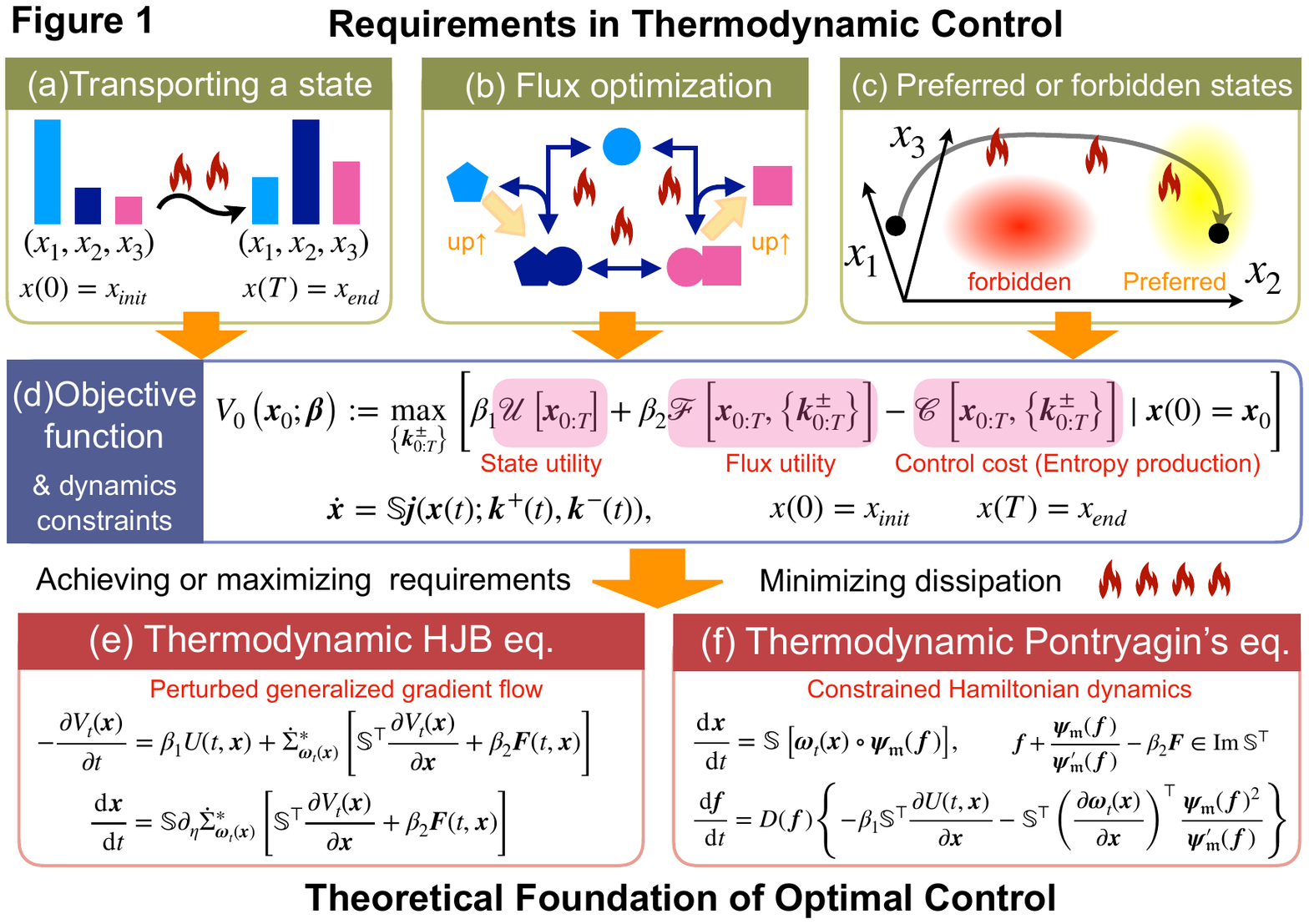}
    \caption{Outline of thermodynamic control theory for MJPs and CRNs. (a) The primary task of thermodynamic control is to drive the state of the system from $\Vc{x}_{init}$ at $t=0$ to $\Vc{x}_{end}$ at $t=T$. Here, the state represents the probability distribution over discrete states for MJPs and the concentration of molecular species for CRNs. (b) The induction of specific transitions (MJPs) or reaction fluxes (CRNs) is indispensable for converting energy into functions. (c) In practical situations, additional constraints on desirable or undesirable states must be accounted for when designing control. (d) Optimal control theory accommodates all these tasks and requirements, along with the thermodynamic cost of control (entropy production or heat dissipation), into a single objective function and minimizes it to obtain the optimal control. (e) The thermodynamic Hamilton-Jacobi-Bellman (HJB) equation characterizes the value of control via the value function, the gradient of which induces the perturbed generalized gradient flow dynamics as the optimally controlled dynamics. (f) The thermodynamic Pontryagin's equation characterizes the optimally controlled Hamiltonian dynamics, which is equivalent to the dynamics obtained by the thermodynamic HJB equation. Here, the equation for $\Vc{F}=\mathrm{const.}$ is shown for notational simplicity. $D(\Vc{\tf})\defeq-\diag\left[\Vc{2}-\frac{\psi_{\gmean}(\Vc{\tf})\psi''_{\gmean}(\Vc{\tf})}{(\psi'_{\gmean}(\Vc{\tf}))^{2}}\right]^{-1}$.}
    \label{fg:outline}
\end{figure}

In this work, we fill this gap by establishing \emph{a thermodynamic optimal-control theory} that is consistent with the thermodynamic and geometric structures of MJPs and CRNs. We consider a class of control problems in which reaction rate constants serve as control variables \cite{horiguchi2025PRXLife}. In addition to an entropy-production-based control cost, we introduce state and flux utility functions into the objective functional (Fig. \ref{fg:outline} (d)). This formulation allows one to treat not only finite-time state transport but also a broader class of control tasks where the occupation of desired states and induction of specified reaction fluxes are balanced against the thermodynamic control cost.

By deriving the thermodynamic version of the Hamilton--Jacobi-Bellman (HJB) equation (Fig. \ref{fg:outline} (e)), we show that the optimal dynamics is a perturbed generalized-gradient flow, with the value function as the driving potential. This result is obtained through a convex representation of the entropy-production rate (EPR) function based on generalized means, and the generalized gradient flow is induced by the convex conjugate of this EPR function as the dissipation function (a nonlinear generalization of the friction metric) \cite{kobayashi2024Info.Geo.}. The derived dynamics has the same thermodynamic structure as the equilibrium and nonequilibrium chemical reaction dynamics\cite{sughiyama2022Phys.Rev.Researcha,kobayashi2022Phys.Rev.Researcha,kobayashi2022Phys.Rev.Research}, which admits its information-geometric characterization\cite{kobayashi2024Info.Geo.}.

We further derive a thermodynamic Pontryagin equation by reframing the Pontryagin's Hamiltonian equation in terms of thermodynamic forces as control variables (Fig. \ref{fg:outline} (f)). This equation manifests the structure specific to thermodynamic control and clarifies its relation to results from optimal transport and related approaches. 
We obtained analytic optimal control solutions for constant activity and a constrained activity budget, which unify the thermodynamic optimal transport solution and steady flux maximization.
Our results establish a theoretical basis for thermodynamic control, with possible applications to biological reaction systems and finite-time thermodynamic operations.

\subsection{Structure of this paper}
The remainder of this paper is organized as follows. Section \ref{sec:Preliminary} reviews the thermodynamic structure of MJPs and CRNs and formulates the optimal control problem in a general setting. In Sec. \ref{sec:Therm_rep_opt_cont}, we derive the thermodynamic HJB and Pontryagin equations by employing a convex representation of the EPR function; this formulation uncovers a perturbed generalized gradient flow structure along with the intrinsic information-geometric features of thermodynamic optimal control problems. 
Building on these results, Sec. \ref{sec:analytic} presents analytical solutions that unify thermodynamic transport and steady-flux maximization.  Finally, Sec. \ref{sec:conclusion} concludes the paper with a discussion of future directions.

\section{Preliminary: Reversible deterministic chemical reaction network and its optimal control}\label{sec:Preliminary}
\subsection{Reversible chemical reaction networks}
Let $\setX\defeq\{X_{1},\cdots, X_{|\setX|}\}$ be the molecular species involved in the system. 
To investigate thermodynamic control, we consider reversible CRN, each pair of forward and backward reactions are defined as 
\begin{align}
\sum_{i=1}^{|\setX|}\gamma^{+}_{i,r}X_{i}\rightleftarrows \sum_{i=1}^{|\setX|}\gamma^{-}_{i,r}X_{i} \label{eq:CRN}
\end{align}
where $r\in[1,|\setR|]$ is the index of reaction.
Here, $\gamma^{+}_{i,r}$ and $\gamma^{-}_{i,r}$ denote the number of $X_i$ molecules consumed and produced in the forward reaction (left to right), respectively. In the reverse reaction (right to left), these roles are inverted, with $\gamma^{-}_{i,r}$ and $\gamma^{+}_{i,r}$ representing the reactants and products.
We define the forward and backward stoichiometric matrices $\stoiMatrix^{\pm} \in \Integer_{\ge 0}^{|\setX|\times |\setR|}$ as $\stoiMatrix^{\pm}_{i,r}\defeq \gamma^{\pm}_{i,r}$.
Then, the standard stoichiometric matrix is defined as $\stoiMatrix\defeq \stoiMatrix^{-}-\stoiMatrix^{+} \in \Integer^{|\setX|\times |\setR|}$.

The deterministic CRN dynamics for the concentration vector $\Vc{x}\in \Real_{\ge0}^{|\setX|}$ of molecular species are described by the chemical rate equation (CRE): 
\begin{align}
\frac{\dd \Vc{x}}{\dt}= \stoiMatrix \left[\Vc{\flux}^{+}(t,\Vc{x})-\Vc{\flux}^{-}(t,\Vc{x})\right]=\stoiMatrix \Vc{\flux}(t,\Vc{x})\label{eq:CRE}
\end{align}
where $\Vc{\flux}^{\pm}(t,\Vc{x})\in \Real_{\ge 0}^{|\setR|}$ are forward and reverse reaction fluxes and their difference $\Vc{\flux}(t,\Vc{x})\defeq \Vc{\flux}^{+}(t,\Vc{x})-\Vc{\flux}^{-}(t,\Vc{x})$ is the (total) thermodynamic flux. In this work, we consider the case that the trajectory is confined within $\Vc{x}\in \Real_{>0}^{|\setX|}$\footnote{More specifically, we exclude the situation that either controlled or uncontrolled trajectories touch the boundary of state space $\Real_{\ge0}^{|\setX|}$.}.
Given the local detailed balance assumption\cite{rao2016Phys.Rev.Xa,maes2021SciPostPhys.Lect.Notesb}, thermodynamic force $\Vc{\tf}$ is defined by 
\begin{align}
\Vc{\tf}(t,\Vc{x})\defeq \ln \Vc{\flux}^{+}(t,\Vc{x})-\ln\Vc{\flux}^{-}(t,\Vc{x}).\label{eq:tf}
\end{align}
The total entropy production rate (EPR) at time $t$ and state $\Vc{x}$ is obtained by 
\begin{align}
\frac{\EPR(t,\Vc{x})}{\gConst\Volume}&= \langle\Vc{\flux}(t,\Vc{x}),\Vc{\tf}(t,\Vc{x}) \rangle =\sum_{r=1}^{|\setR|}\left(\flux_{r}^{+}(t,\Vc{x})-\flux_{r}^{-}(t,\Vc{x})\right)\ln \frac{\flux_{r}^{+}(t,\Vc{x})}{\flux_{r}^{-}(t,\Vc{x})}\\
&=\KLD_{KL}[\Vc{\flux}^{+}(t,\Vc{x})\|\Vc{\flux}^{-}(t,\Vc{x})]+\KLD_{KL}[\Vc{\flux}^{-}(t,\Vc{x})\|\Vc{\flux}^{+}(t,\Vc{x})]\ge 0 \label{eq:epr}
\end{align}
where $R$ is the gas constant, $\Volume$ is the volume, $\langle\Vc{\tf},\Vc{\flux}\rangle\defeq \sum_{r=1}^{|\setR|}\tf_{r}\flux_{r}$ is the bilinear pairing, and  $\KLD_{KL}[\Vc{\flux}\|\Vc{\flux}']\defeq \sum_{r=1}^{|\setR|}\left[\flux_{r}\ln \frac{\flux_{r}}{\flux'_{r}}-(\flux_{r}-\flux'_{r})\right]$ is the Kullbuck-Leibler divergence for non-negative densities. In the following, we set $\gConst\Volume=1$ for simplicity.

\subsection{Mass Action Kinetics}
While the definitions of thermodynamic force and EPR are independent of the functional forms of the fluxes with respect to $\Vc{x}$, specific kinetics for the forward and backward fluxes are required to instantiate the general theory that we derive.  
Mass action kinetics (MA kinetics) is the representative one. 
For the reversible CRN(\eqnref{eq:CRN}), their forward and backward flux functions under MA kinetics are 
\begin{align}
\flux^{\pm}_{r}(\Vc{x})&\defeq \kcoef^{\pm}_{r}\Vc{x}^{(\Vc{\gamma}^{\pm}_{r})^{\Transpose}}=\kcoef^{\pm}_{r}\prod_{i=1}^{|\setX|}x_{i}^{\gamma^{\pm}_{i,r}} \label{eq:MA_kinetics}
\end{align}
where $\Vc{\kcoef}^{\pm}\defeq (\kcoef_{1}^{\pm},\cdots,\kcoef_{|\setR|}^{\pm})^{\Transpose}\in \Real_{\ge 0}^{|\setR|}$ are the kinetic rate constants and $\Vc{x}^{(\Vc{\gamma}_{r})^{\Transpose}}$ is a short-hand notation of $\prod_{i=1}^{|\setX|}x_{i}^{\gamma_{i,r}}$ (see also the Appendix for the detailed setup.). 
In addition, if all reactions are monomolecular and thus their fluxes are linear, the CRE (\eqnref{eq:CRE}) is reduced to the master equation for MJP on a graph. The flux functions are reduced to
\begin{align}
\flux^{\pm}_{r}(\Vc{x})\defeq \kcoef^{\pm}_{r}\Vc{x}^{(\Vc{b}_{r}^{\pm})^{\Transpose}}=\kcoef^{\pm}_{r}\prod_{i=1}^{|\setX|}x_{i}^{b_{i,r}^{\pm}}\overset{(*)}{=}\kcoef^{\pm}_{r}\left[\sum_{i=1}^{|\setX|}b_{i,r}^{\pm}x_{i}\right]=\kcoef^{\pm}_{r}(\Vc{b}_{r}^{\pm})^{\Transpose}\Vc{x} \label{eq:MA_kinetics_MJP}
\end{align}
where $b_{i,r}^{\pm}\in\{0,1\}$ is $1$ if the $i$th molecule is the reactant (product) of the $r$th reaction, respectively, and $0$ otherwise. 
For $\indMatrix^{\pm}\defeq\{b_{i,r}\}$, $\indMatrix=\indMatrix^{+}-\indMatrix^{-}\in \{-1,0,+1\}^{|\setX|\times|\setR|}$ is the graph incidence matrix for the MJP\footnote{In this case, the stoichiometric matrix is related with the graph incidence matrix as $-\stoiMatrix=\indMatrix\defeq (b_{i,r})$.}. $R\Volume$ in \eqnref{eq:epr} is replaced with $k_{B}$.

\subsection{Optimal control of deterministic CRNs}
To introduce external control to a CRN, we decompose the forward and backward fluxes as 
\begin{align}
\Vc{\flux}^{\pm}(t,\Vc{x})=\Vc{\kcoef}^{\pm}(t)\circ\Vc{h}^{\pm}(\Vc{x}),
\end{align}
where $\Vc{\kcoef}^{\pm}(t) \ge \Vc{0}$ are the reaction rate constants and $\Vc{h}^{\pm}(\Vc{x})\ge \Vc{0}$ are the $\Vc{x}$ dependence of the flux functions that satisifes $\Vc{h}^{\pm}(\Vc{x})> \Vc{0}$ if $\Vc{x}\in \Real_{>0}^{|\setX|}$.
We suppose that we can control the reaction rate constants $\Vc{\kcoef}^{\pm}(t)$ independently. 
Under this setup, we consider the EPR $\EPR(\Vc{x},\{\Vc{\kcoef}^{\pm}\})$ as the cost for controlling the kinetic rate of a CRN to $\{\Vc{\kcoef}^{\pm}\}\defeq (\Vc{\kcoef}^{+}, \Vc{\kcoef}^{-})$ at state $\Vc{x}$:
\begin{align}
\costF(\Vc{x},\{\Vc{\kcoef}^{\pm}\})\defeq \EPR(\Vc{x},\{\Vc{\kcoef}^{\pm}\}).
\end{align}
Then, we introduce the control cost functional over the trajectory $\Vc{x}_{0:T}\defeq\{\Vc{x}_{t}|t\in[0,T]\}$ and $\Vc{\kcoef}^{\pm}_{0:T}\defeq\{\Vc{\kcoef}^{\pm}_{t}|t\in[0,T]\}$ as
\begin{align}
    \costF^{\pm}[\Vc{x}_{0:T},\{\Vc{\kcoef}^{\pm}_{0:T}\}] 
    &\defeq \int_{0}^{T}\costF(\Vc{x}_{t},\{\Vc{\kcoef}^{\pm}_{t}\})\dt=
    \int_{0}^{T}\EPR(\Vc{x}_{t},\{\Vc{\kcoef}^{\pm}_{t}\})\dt.
\end{align}

Conventional thermodynamic transport problems only account for this control cost. 
However, in various practical situations, we may have additional constraints and requirements for the operation of the thermodynamic system\cite{brown2017Proc.Natl.Acad.Sci.,andrieux2023,yang2026NatCommun,brown2017Proc.Natl.Acad.Sci.,andrieux2023,yang2026NatCommun,schmiedl2007EPL,fleming2012JournalofTheoreticalBiology,gagrani2026J.Stat.Mech.,pinero2024Phys.Rev.Res.,bilancioni2025NatCommun,horowitz2017Phys.Rev.E}.
To integrate these requirements, we introduce the instantaneous utility function $U(t,\Vc{x})\in \Real$, the terminal utility function $\hat{U}_{T}(\Vc{x})\in \Real$ and the flux utility function $\Vc{F}^{\pm}(t,\Vc{x})=(F_{1}^{\pm}(t,\Vc{x}),\cdots,F_{|\setR|}^{\pm}(t,\Vc{x}))^{\Transpose}\in \Real^{|\setR|}$. The instantaneous utility function $U(t,\Vc{x})$ specifies the preferred state of $\Vc{x}$ at time $t$ (Fig. \ref{fg:outline} (c)), whereas the terminal utility determines the state where the system is encouraged to reach at the end of control. If the terminal state $\Vc{x}_{end}$ is specified, then $\hat{U}_{T}(\Vc{x})=0$ if $\Vc{x}=\Vc{x}_{end}$ and $\hat{U}_{T}(\Vc{x})=-\infty$, otherwise.
The $r$th component of the flux utilities,  $F_{r}^{+}(t,\Vc{x})$ and $F_{r}^{-}(t,\Vc{x})$, define the desiability of the occurence of the $r$th forward and backward reactions, respectively (Fig. \ref{fg:outline} (b)). In this work, we assume that the flux utilities for each reversible reaction are anti-symmetric (reversible with respect to flux), i.e.:
$F_{r}^{+}(t,\Vc{x})=-F_{r}^{-}(t,\Vc{x})\eqdef F_{r}(t,\Vc{x})$.
For these utilities, we introduce their integral over the trajectory:
\begin{align}
    \sutilF[\Vc{x}_{0:T}] &\defeq \int_{0}^{T}U(t,\Vc{x}(t))  \dt + \hat{U}_{T}(\Vc{x}(T)) \\ 
    \futilF[\Vc{x}_{0:T}, \{\Vc{\kcoef}^{\pm}_{0:T}\}] 
    &\defeq \int_{0}^{T}\sum_{\mathfrak{s}\in\{-,+\}}\sum_{r}F^{\mathfrak{s}}_{r}(t,\Vc{x}(t))   \flux^{\mathfrak{s}}_{r}(t)\dt=\int_{0}^{T}\left\langle \Vc{F}(t,\Vc{x}(t)),  \Vc{\flux}(t)\right\rangle \dt     
\end{align}
where we used the anti-symmetry of flux utilities and abbreviate the dependence of the flux on $\{\Vc{\kcoef}^{\pm}(t)\}$ as $\Vc{\flux}(t)=\Vc{\flux}(\Vc{x}(t);\{\Vc{\kcoef}^{\pm}(t)\})$.

Then, the optimal control problem for a finite horizon $T$ starting from $\Vc{x}_{init}$ is defined as 
\begin{align}
    \valueF_{0}(\Vc{x}_{0}; \Vc{\beta})&\defeq \max_{\{\Vc{\kcoef}^{\pm}_{0:T}\}}\left[\beta_{1}\sutilF[\Vc{x}_{0:T}] + \beta_{2}\futilF[\Vc{x}_{0:T}, \{\Vc{\kcoef}^{\pm}_{0:T}\}] - \costF[\Vc{x}_{0:T}, \{\Vc{\kcoef}^{\pm}_{0:T}\}]\mid \Vc{x}(0)=\Vc{x}_{init}\right]\notag \\
    &\mathrm{s.t.} \quad \frac{\dd \Vc{x}}{\dt}=\stoiMatrix\Vc{\flux}(\Vc{x}(t);\{\Vc{\kcoef}^{\pm}(t)\}),\label{eq:valueF_def}
\end{align}
where $\Vc{\beta}=(\beta_{1},\beta_{2})$ is the weighting parameters (Fig. \ref{fg:outline} (d)).

\subsection{Bellman's Principle of Optimality and HJB equation}
A standard approach for solving this control problem is to follow Bellman Principle of Optimality, which decomposes the optimization problem over $t'\in[t,T]$ into one at $t$ and the sub-problem over $t'\in[t+\dt,T]$. To conduct this decomposition operationally, we introduce the value function:
\begin{align}
    \valueF_{t}(\Vc{x}; \Vc{\beta})&\defeq \max_{\{\Vc{\kcoef}^{\pm}_{t:T}\}}\left[\beta_{1}\sutilF[\Vc{x}_{t:T}] + \beta_{2} \futilF[\Vc{x}_{t:T}, \{\Vc{\kcoef}^{\pm}_{t:T}\}] - \costF[\Vc{x}_{t:T}, \{\Vc{\kcoef}^{\pm}_{t:T}\}]\mid \Vc{x}(t)=\Vc{x}\right],\label{eq:v_func_path}
\end{align}
which satisfies the Hamilton Jacobi Bellman (HJB) equation:
\begin{align}
    -&\frac{\partial \valueF_{t}(\Vc{x};\Vc{\beta})}{\partial t}=\beta_{1}U(t,\Vc{x}) \notag \\
    &+ \max_{\{\Vc{\kcoef}^{\pm}\}\in \Real_{\ge 0}^{2 |\setR|}}\left[ \left\langle \frac{\partial \valueF_{t}(\Vc{x};\Vc{\beta})}{\partial \Vc{x}}, \frac{\dd \Vc{x}}{\dt}\right\rangle+\beta_{2}\langle\Vc{F}(t,\Vc{x}) ,\Vc{\flux}(\Vc{x}; \{\Vc{\kcoef}^{\pm}\})\rangle 
    - \costF(\Vc{x},\{\Vc{\kcoef}^{\pm}\})\right] \label{eq:HJB_orig_det},
\end{align}
where the terminal boundary condition is $\valueF_{T}(\Vc{x};\Vc{\beta})=\hat{U}_{T}(\Vc{x})$.
In the HJB equation, the optimization over the trajectory is replaced by the optimization at $t$ given $\valueF_{t}(\Vc{x};\Vc{\beta})$.
Thus, the optimal $\Vc{\kcoef}^{\pm}_{t}$ becomes 
\begin{align}
\Vc{\kcoefop}^{\pm}_{t}(\Vc{x};\Vc{\beta}) &\defeq \arg\max_{\{\Vc{\kcoef}^{\pm}\}\in \Real_{\ge 0}^{2 |\setR|}}\left[ \left\langle \frac{\partial \valueF_{t}(\Vc{x};\Vc{\beta})}{\partial \Vc{x}}, \frac{\dd \Vc{x}}{\dt}\right\rangle+\beta_{2}\langle\Vc{F}(t,\Vc{x}) ,\Vc{\flux}(\Vc{x}; \{\Vc{\kcoef}^{\pm}\})\rangle 
    - \costF(\Vc{x},\{\Vc{\kcoef}^{\pm}\})\right]\\
&= \arg\max_{\{\Vc{\kcoef}^{\pm}\}\in \Real_{\ge 0}^{2 |\setR|}}\left[ \left\langle \left(\stoiMatrix^{\Transpose}\frac{\partial \valueF_{t}(\Vc{x};\Vc{\beta})}{\partial \Vc{x}}+\beta_{2}\Vc{F}(t,\Vc{x})\right),\Vc{\flux}(\Vc{x};\{\Vc{\kcoef}^{\pm}\})\right\rangle 
    - \costF(\Vc{x},\{\Vc{\kcoef}^{\pm}\})\right]\label{eq:opt_k}
\end{align}
where we used $\frac{\dd \Vc{x}}{\dt}=\stoiMatrix \Vc{\flux}(\Vc{x};\{\Vc{\kcoef}^{\pm}\})$.
If an explicit form of $\Vc{\kcoefop}^{\pm}_{t}$ is provided, the value function can be obtained by solving 
\begin{align}
    -&\frac{\partial \valueF_{t}(\Vc{x};\Vc{\beta})}{\partial t}=\beta_{1}U(t,\Vc{x}) +\left\langle \left(\stoiMatrix^{\Transpose}\frac{\partial \valueF_{t}(\Vc{x};\Vc{\beta})}{\partial \Vc{x}}+\beta_{2}\Vc{F}(t,\Vc{x})\right),\Vc{\flux}(\Vc{x};\{\Vc{\kcoefop}^{\pm}_{t}(\Vc{x})\})\right\rangle 
    - \costF(\Vc{x},\{\Vc{\kcoefop}^{\pm}_{t}(\Vc{x})\})
    \label{eq:HJB_orig_det_after_opt},
\end{align}
in a time-backward manner from the terminal boundary condition $\valueF_{T}(\Vc{x};\Vc{\beta})=\hat{U}_{T}(\Vc{x})$.

\section{Thermodynamic representations of optimal control for CRNs}\label{sec:Therm_rep_opt_cont}
Solving \eqnref{eq:opt_k} explicitly is intractable for general control problems with nonlinearities in the cost and utility functions. 
If the cost function is strictly convex with respect to the control variable, the problem becomes manageable at least formally by using the Legendre transformation. However, the thermodynamic cost $\costF(\Vc{x},\{\Vc{\kcoef}^{\pm}_{t}(\Vc{x})\})=\EPR$ is convex but not strict with respect to $\Vc{\kcoef}^{\pm}$ due to a degeneracy\footnote{Specifically, $\EPR(\lambda \Vc{\kcoef}^{+}, \lambda \Vc{\kcoef}^{-})$ is linear with respect to $\lambda>0$.}.
We resolve this degeneracy by applying transformations based on generalized means\cite{nagayama2025Phys.Rev.Res.}, which decouples the degeneracy and the strict convexity.

\subsection{Generalized mean and thermodynamic representation of flux}

For two non-negative scalars $a, b\ge 0$, let us consider a class of their generalized mean $\gmean[a,b]\ge0$, which satisfies $\min[a,b] \le \gmean[a,b]\le \max[a,b]$, equality $\gmean[a,a]=a$, symmetry $\gmean[a,b]=\gmean[b,a]$, and homogeneity $ \gmean(\lambda a,\lambda b)=\lambda \gmean[a,b]$ for any $\lambda>0$. By defining $\gmean(r)\defeq \gmean[r,1]$, we can represent $\gmean[a,b]=b\, \gmean(a/b)$\footnote{Note that $\tf \cdot \gmean(1/\tf)=\gmean(\tf)$ holds from the symmetry. In addition, for a differentiable $\gmean(r)$, we have $\gmean(r)-r\gmean'(r)= \gmean'(1/r)$ where $\gmean'(r)\defeq \frac{\dd \gmean(r)}{\dd r}$. See \cite{nagayama2025Phys.Rev.Res.} for more detail.}.
With $\gmean(r)$, the generalized average flux is introduced\cite{nagayama2025Phys.Rev.Res.} as 
\begin{align}
\Vc{\omega}\defeq \gmean[\Vc{\flux}^{+}, \Vc{\flux}^{-}]\in \Real_{\ge 0}^{|\setR|} \label{eq:gmean}
\end{align}
where $\omega_{r}=\gmean[\flux^{+}_{r}, \flux^{-}_{r}]$.
This quantity (up to constant scaling)  is called ``traffic'' for the arithmetic mean\cite{yoshimura2023Phys.Rev.Res.a}, ``Onsager coefficients'' or ``state mobility'' for logarithmic mean\cite{yoshimura2023Phys.Rev.Res.a,nagayama2025Phys.Rev.Res.}, and ``frenetic activity'' for geometric mean\cite{maes2018,kobayashi2022Phys.Rev.Research}. Here we call it "general activity" or simply "activity" as in \cite{nagayama2025Phys.Rev.Res.}.
Then, the total flux can be represented as 
\begin{align}
    \Vc{\flux}\defeq \Vc{\flux}^{+}-\Vc{\flux}^{-}=\Vc{\omega}\circ \Vc{\psi}_{\gmean}(\Vc{\tf}) \label{eq:j_f}
\end{align}
    where $\Vc{\psi}_{\gmean}(\Vc{\tf})=(\psi_{\gmean}(\tf_{1}), \cdots, \psi_{\gmean}(\tf_{|\setR|}))^{\Transpose}$ with
\begin{align}
\psi_{\gmean}(\tf)\defeq \frac{e^{\tf}-1}{\gmean(e^{\tf})}\in \Real.
\end{align}
The $\gmean$-induced function $\psi_{\gmean}(\tf)$ satisfies oddity $\psi_{\gmean}(\tf)=-\psi_{\gmean}(-\tf)$, which leads to $\psi_{\gmean}(\tf)=0$ if $\tf=0$.
Given $\Vc{\omega}$ and $\Vc{\tf}\neq \Vc{0}$, the corresponding $\Vc{\flux}^{\pm}$ are recovered as $\Vc{\flux}^{+}=\Vc{\omega}\circ\frac{ \Vc{\psi}_{\gmean}(\Vc{\tf})}{\Vc{1}-\exp(-\Vc{\tf})}$, $\Vc{\flux}^{-}=\Vc{\omega}\circ\frac{ \Vc{\psi}_{\gmean}(\Vc{\tf})}{\exp(\Vc{\tf})-1}$ for $\Vc{\tf}\neq 0$, and $\Vc{\flux}^{+}=\Vc{\flux}^{-}=\Vc{\omega}$ for $\Vc{\tf}= 0$\footnote{It should be noted that $\lim_{\Vc{\tf}\to\Vc{0}}\frac{ \psi_{\gmean}(\tf)}{1-\exp(-\tf)}=1$.}.
In this work, we consider only the differentiable generalized means that satisfy the two physical conditions
\begin{align}
\gmean'(r)+\gmean'(1/r)&>0 & \mbox{and }&&  \gmean''(r)&> -\frac{r+1}{r(r-1)^{2}}\left[\gmean'(r)+\gmean'(1/r)\right] \label{eq:phys_cond}
\end{align}
for $r >1$ as in\cite{nagayama2025Phys.Rev.Res.}.
Examples of $\gmean$ and $\psi_{\gmean}$ are listed in Table \ref{tb:means}, all of which satisfy these physical conditions.
\begin{table}[h]
\caption{A list of generalized means and the corresponding functional forms of $\psi_{\gmean}(\tf)$ and $\psi_{\gmean}^{-1}(u)$ with their domains for $u$. The domin for $\tf$ is $D^{f}=(-\infty, +\infty)$ for all means. 
}
\centering
\begin{tabular}{|l|c|c|c|c|c|}
\hline
\textbf{Type of mean}
& $\gmean[\flux^{+},\flux^{-}]$
& $\gmean(r)$
& $\psi_{\gmean}(\tf)$
& $\psi_{\gmean}^{-1}(u)$
& \textbf{Domain} $D^{u}$
\\
\hline
\hline
Harmonic & $\dfrac{2\flux^{+}\flux^{-}}{\flux^{+}+\flux^{-}}$ & $\dfrac{2 r}{r+1}$ & $\sinh(\tf)$ & $\sinh^{-1}(u)$
& $u\in (-\infty, +\infty)$\\
\hline
Geometric & $\sqrt{\flux^{+}\flux^{-}}$ & $\sqrt{r}$ & $2\sinh(\tf/2)$ & $2\sinh^{-1}(u/2)$ 
& $u\in (-\infty, +\infty)$\\
\hline
Logarithmic & $\dfrac{\flux^{+}-\flux^{-}}{\ln \flux^{+}-\ln \flux^{-}}$ & $\dfrac{r-1}{\ln(r)}$ & $\tf$ & $u$ 
& $u\in (-\infty, +\infty)$\\
\hline
Arithmetic & $\dfrac{\flux^{+}+\flux^{-}}{2}$ & $\dfrac{r+1}{2}$ & $2\tanh(\tf/2)$ & $2\tanh^{-1}(u/2)$ 
& $u\in (-2,+2)$\\
\hline
Contraharmonic & $\dfrac{(\flux^{+})^{2}+(\flux^{-})^{2}}{\flux^{+}+\flux^{-}}$ & $\dfrac{r^{2}+1}{r+1}$ & $\tanh(f)$ & $\tanh^{-1}(u)$ 
& $u\in (-1,+1)$\\
\hline
\end{tabular}
\label{tb:means}
\end{table}

\begin{figure}
    \centering
    \includegraphics[width=0.95\linewidth]{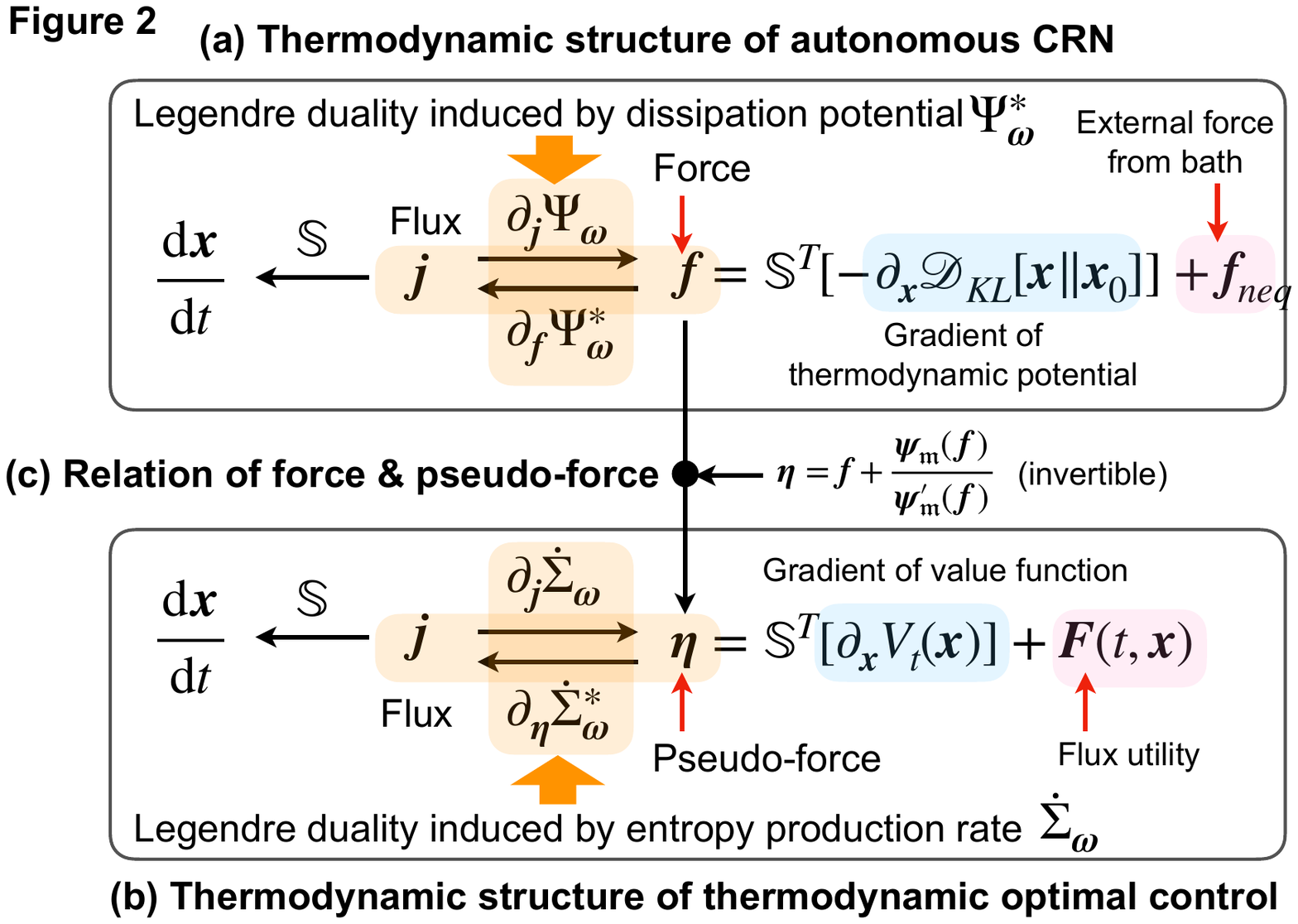}
    \caption{The thermodynamic structure of autonomous CRNs (a) and the optimally controlled CRNs (b). Both dynamics are the perturbed generalized gradient flow. For the autonomous CRNs (a), the dynamics is induced by the thermodynamic force generated by the gradient of the thermodynamic potential function $\KLD[\Vc{x}\|\Vc{x}_{eq}]$ perturbed by the external nonequilibrium force $\Vc{f}_{ne}$. The flux $\Vc{\flux}$ is the Legendre transformation of the force $\Vc{\tf}$ via the dissipation potential $\Psi^{*}_{\Vc{\omega}(\Vc{x})}(\Vc{\tf})$. For the optimally controlled CRNs (b), the dynamics is induced by the pseudo-force $\Vc{\tff}$ generated by the gradient of the value function $\valueF_{t}(\Vc{x})$ perturbed by the flux utility $\Vc{F}=\Vc{F}(t,\Vc{x})$. The flux $\Vc{\flux}$ is the Legendre transformation of the pseudo-force via the Legendre dual of the EPR $\EPR^{*}_{\Vc{\omega}(\Vc{x})}(\Vc{\tff})$. }
    \label{fig:CRN_control}
\end{figure}

When the first physical condition in \eqnref{eq:phys_cond} is satisfied, $u=\psi_{\gmean}(\tf)$ becomes strictly monotonous and thus invertible $\tf=\psi_{\gmean}^{-1}(u)$. 
Then, give $\Vc{\omega}$ and $\Vc{\flux}$, the corresponding $\Vc{\flux}^{\pm}$ are recovered as $\Vc{\flux}^{\pm}=\pm\frac{\Vc{\flux}}{\Vc{1}-\exp(\mp\psi_{\gmean}^{-1}(\Vc{\flux}/\Vc{\omega})}$ if $\Vc{u}=\Vc{\flux}/\Vc{\omega}$ is in the corresponding domain.
Furthermore, by defining dissipation functions,  $\dissip_{\Vc{\omega}}(\Vc{\flux}), \dissip^{*}_{\Vc{\omega}}(\Vc{\tf})\in \Real_{\ge 0}$ as
\begin{align}
\dissip_{\Vc{\omega}}(\Vc{\flux})&=\sum_{r=1}^{|\setR|}\omega_{r}\int_{\Vc{0}}^{|\flux_{r}|/\omega_{r}}\psi_{\gmean}^{-1}(u')\dd u' &
\dissip^{*}_{\Vc{\omega}}(\Vc{\tf})&=\sum_{r=1}^{|\setR|}\omega_{r}\int_{\Vc{0}}^{|\tf_{r}|}\psi_{\gmean}(\tf')\dd \tf',
\end{align}
they are strictly convex, symmetric, and convex conjugate to each other. The thermodynamic force and flux are associated as the Legendre dual of these functions:
\begin{align}
\Vc{\flux}&=\partial_{\Vc{\tf}} \dissip^{*}_{\Vc{\omega}}[\Vc{\tf}], &\Vc{\tf}&=\partial_{\Vc{\flux}}\dissip_{\Vc{\omega}}[\Vc{\flux}] ,\label{eq:dual_j_f}
\end{align}
where $\partial_{\Vc{\tf}}\defeq \frac{\partial}{\partial \Vc{\tf}}$ and $\partial_{\Vc{\flux}}\defeq \frac{\partial}{\partial \Vc{\flux}}$.
Owing to the Fenchel--Young inequality, EPR (\eqnref{eq:epr}) is represented as 
\begin{align}
\EPR=\langle\Vc{\flux},\Vc{\tf} \rangle = \dissip_{\Vc{\omega}}(\Vc{\flux}) + \dissip^{*}_{\Vc{\omega}}(\Vc{\tf}) \label{eq:EPR_var}
\end{align}
Moreover, \eqnref{eq:CRE} with the MA kinetics is represented as the perturbed generalized gradient flow (Fig. \ref{fig:CRN_control} (a))\footnote{This representation can be generalized by replacing the KL divergence with the Bregman divergence $\KLD_{\Phi}(\Vc{x}\|\Vc{x}_{eq})\defeq \Phi(\Vc{x})-\Phi(\Vc{x}_{eq})-\langle\partial \Phi(\Vc{x}_{eq})/\partial\Vc{x},\Vc{x}-\Vc{x}_{eq} \rangle$ induced by the thermodynamic potential function $\Phi(\Vc{x})$. See \cite{kobayashi2024Info.Geo.} for the detail.}:
\begin{align}
\frac{\dd \Vc{x}}{\dt}=\stoiMatrix\Vc{\flux}(\Vc{x})=\stoiMatrix \partial_{\Vc{\tf}} \dissip^{*}_{\Vc{\omega}(\Vc{x})}[\Vc{\tf}(\Vc{x})]=\stoiMatrix \partial_{\Vc{\tf}} \dissip^{*}_{\Vc{\omega}(\Vc{x})}\left[-\stoiMatrix^{\Transpose} \frac{\partial\KLD_{KL}(\Vc{x}\|\Vc{x}_{eq})}{\partial \Vc{x}}+\Vc{f}_{ne}\right] \label{eq:CRN_var}.
\end{align}
 
\subsection{Convex representation of entropy production rate by generalized mean}
\begin{figure}
    \centering    \includegraphics[width=0.95\linewidth]{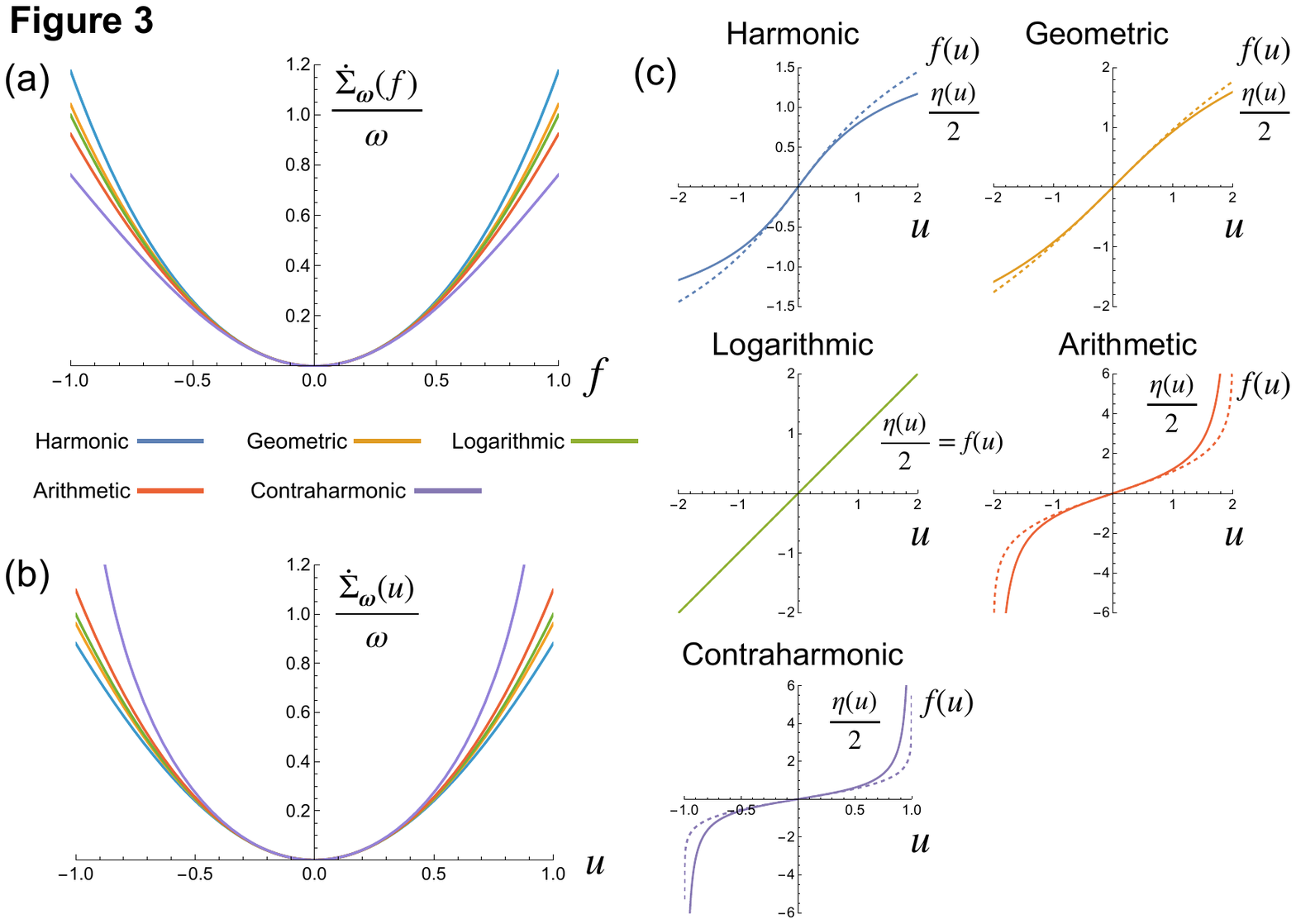}
    \caption{The functional form of EPR as a function of $\tf$ (a) and $u=\flux/\omega$ (b) for each generalized mean where $|\setR|=1$. (c) Comparison of the pseudo-force $\tff$ (the solid curve) and the thermodynamic force $\tf$ (the dashed curve) as function of $u=\flux/\omega$ for each generalized mean.}
    \label{fig:EPR_f_eta}
\end{figure}

While the representation \eqnref{eq:EPR_var} of EPR by the pair of convex dissipation functions is fundamental in the thermodynamics of CRNs, it may not be employed for the control problem because it is a sum of two strictly convex functions with different arguments, i.e., $\Vc{\flux}$ and $\Vc{\tf}$.
 The EPR is more directly represented as functions of $\Vc{\tf}$, $\Vc{\flux}$, or $\Vc{u}$ by using \eqnref{eq:j_f} and its inverse $\Vc{\tf}= \Vc{\psi}^{-1}_{m}\left(\frac{\Vc{\flux}}{\Vc{\omega}}\right)$ (Fig. \ref{fig:EPR_f_eta} (a) and (b)): 
\begin{align}
\EPR_{\Vc{\omega}}(\Vc{\tf})&=\langle \Vc{\omega}, \Vc{f}\circ\Vc{\psi}_{m}(\Vc{f})\rangle \label{eq:EPR_f}\\
\EPR_{\Vc{\omega}}(\Vc{\flux})&=\left\langle  \Vc{\omega}, \frac{\Vc{\flux}}{\Vc{\omega}}\circ \Vc{\psi}_{m}^{-1}\left(\frac{\Vc{\flux}}{\Vc{\omega}}\right)\right\rangle \label{eq:EPR_j}\\
\EPR_{\Vc{\omega}}(\Vc{u})&= \left\langle  \Vc{\omega}, \Vc{u}\circ \Vc{\psi}_{m}^{-1}\left(\Vc{u}\right)\right\rangle \label{eq:EPR_u}
\end{align}
If the second condition in \eqnref{eq:phys_cond} is  met additionally to the first one\footnote{In terms of $\psi_{\gmean}(\tf)$, this condition is equivalent to $2(\psi'_{\gmean}(\tf))^{2}-\psi_{\gmean}(\tf)\psi''_{\gmean}(\tf)>0$ for all $\Vc{\tf}$.}, $\EPR_{\Vc{\omega}}(\Vc{u})$ becomes strictly convex. Furthermore, $\EPR_{\Vc{\omega}}(\Vc{u})$ is symmetric and greater than or equal to $0$, the equality of which holds if and only if $\Vc{u}=0$. The same true for $\EPR_{\Vc{\omega}}(\Vc{\flux})$\footnote{It should be noted that $\EPR_{\Vc{\omega}}(\Vc{\tf})$ is not necessarily convex. For example, $\tf \tanh(\tf)$, corresponding to the contraharmonic mean, is not convex.}.
Then, the convex conjugate of $\EPR_{\Vc{\omega}}(\Vc{\flux})$ is defined as 
\begin{align}
\EPR^{*}_{\Vc{\omega}}(\Vc{\tff})\defeq \sup_{\Vc{\flux}}\left[\langle\Vc{\tff},\Vc{\flux} \rangle-\EPR_{\Vc{\omega}}(\Vc{\flux})\right],
\end{align}
which is also strictly convex, symmetric, and greater than or equal to $0$ from the properties of the convex conjugate.  
Thus, the pair of $\EPR_{\Vc{\omega}}(\Vc{\flux})$ and $\EPR^{*}_{\Vc{\omega}}(\Vc{\tff})$ induce the Legendre duality between $\Vc{\flux}$ and $\Vc{\tff}$ as 
\begin{align}
\Vc{\flux}&=\partial_{\Vc{\tff}} \EPR^{*}_{\Vc{\omega}}(\Vc{\tff}), & 
\Vc{\tff}&=\partial_{\Vc{\flux}}\EPR_{\Vc{\omega}}(\Vc{\flux}).
\end{align}
We refer to $\Vc{\tff}$, the Legendre dual of $\Vc{\flux}$ via EPR, as the pseudo-force because $\Vc{\tff}$ differs from the thermodynamic force $\Vc{\tf}$ (Fig. \ref{fig:EPR_f_eta} (c)), which is the Legendre dual of $\Vc{\flux}$ induced by the dissipation function (\eqnref {eq:dual_j_f}). 
The explicit representation of $\Vc{\eta}$ as a function of $\Vc{\flux}$ is  $\Vc{\eta}=\Vc{\psi}^{-1}_{\gmean}(\Vc{\flux}/\Vc{\omega})+\frac{\Vc{\flux}}{\Vc{\omega}}\circ   (\Vc{\psi}^{-1}_{\gmean})'(\Vc{\flux}/\Vc{\omega})$ where $(\psi^{-1}_{\gmean})'(\tf)\defeq \dd \psi^{-1}_{\gmean}(\tf)/\dd \tf$. In contrast, $\Vc{\flux}$ and also $\EPR_{\Vc{\omega}}^{*}(\Vc{\tff})$ are not explicitly represented by $\Vc{\eta}$, except for the logarithmic mean.
Nevertheless, as we see in Sec. \ref{sec:ThermoRepPont},  $\EPR^{*}_{\Vc{\omega}}(\Vc{\tff}(\Vc{\flux}))\defeq \left.\EPR^{*}_{\Vc{\omega}}(\Vc{\tff})\right|_{\Vc{\eta}=\partial_{\Vc{}}\EPR_{\Vc{\omega}}(\Vc{\flux})}$, $\EPR_{\Vc{\omega}}^{*}(\Vc{\tff})$ as a function of $\Vc{\flux}$, has an explicit form $\EPR^{*}_{\Vc{\omega}}(\Vc{\tff}(\Vc{\flux}))= \left\langle \Vc{\flux},\frac{\partial^{2}\dissip_{\Vc{\omega}}(\Vc{\flux})}{\partial \Vc{\flux}^{2}}\Vc{\flux} \right\rangle$, which manifests that it is identical to the generalized pseudo-EPR $\Pi_{\Vc{\omega}}(\Vc{\flux})$.

\begin{table}[h]
\caption{The functional forms of the ERP and its conjugate and induced Legendre transformation for representative generalized means. NA stands for no analytic representation by elementary functions available.}
\centering
\begin{tabular}{|l|c|c|c|c|c|}
\hline
\textbf{Type of mean}
&$\boldsymbol{\omega[\flux^{+},\flux^{-}]}$
& $\EPR_{\Vc{\omega}}(\Vc{\flux})$
& $\Vc{\tff}=\partial_{\Vc{\flux}} \EPR_{\Vc{\omega}}(\Vc{\flux})$
& $\EPR_{\Vc{\omega}}^{*}(\Vc{\tff})$
& $\Vc{\flux}=\partial \EPR_{\Vc{\omega}}^{*}(\Vc{\tff})$
\\
\hline
\hline
Harmonic & $\dfrac{2\Vc{\flux}^{+}\circ\Vc{\flux}^{-}}{\Vc{\flux}^{+}+\Vc{\flux}^{-}}$ & $\Vc{\flux}\circ \sinh^{-1}\left(\frac{\Vc{\flux}}{\Vc{\omega}}\right)$
& $\frac{\Vc{\flux}^{2}}{\sqrt{\Vc{\omega}^{2}+\Vc{\flux}^{2}}}+ \sinh^{-1}\left(\frac{\Vc{\flux}}{\Vc{\omega}}\right)$ & \mbox{NA}& \mbox{NA}\\
\hline
Geometric &$\sqrt{\Vc{\flux}^{+}\circ\Vc{\flux}^{-}}$ & $2\Vc{\flux}\circ\sinh^{-1}\left(\frac{\Vc{\flux}}{2\Vc{\omega}}\right)$ 
& $\frac{\Vc{\flux}^{2}}{\sqrt{\Vc{\omega}^{2}+\frac{\Vc{\flux}^{2}}{4}}}+2 \sinh^{-1}\left(\frac{\Vc{\flux}}{2\Vc{\omega}}\right)$& \mbox{NA}& \mbox{NA}\\
\hline
Logarithmic & $\dfrac{\Vc{\flux}^{+}-\Vc{\flux}^{-}}{\ln \Vc{\flux}^{+}-\ln \Vc{\flux}^{-}}$ & $\Vc{\flux}^{2}/\Vc{\omega}$
 &$2\Vc{\flux}/\Vc{\omega}$ & $\Vc{\omega}\circ\left(\frac{\Vc{\tff}}{2}\right)^{2}$
&$\frac{\Vc{\omega}}{2}\circ \Vc{\tff}$\\
\hline
Arithmetic &$\dfrac{\Vc{\flux}^{+}+\Vc{\flux}^{-}}{2}$ & $2\Vc{\flux}\circ\tanh^{-1}\left(\frac{\Vc{\flux}}{2\Vc{\omega}}\right)$ 
& $\frac{\Vc{\omega}\circ \Vc{\flux}}{\Vc{\omega}^{2}-\frac{\Vc{\flux}^{2}}{4}}+2 \tanh^{-1}\left(\frac{\Vc{\flux}}{2\Vc{\omega}}\right)$& \mbox{NA}& \mbox{NA}\\
\hline
Contraharmonic &$\dfrac{(\Vc{\flux}^{+})^{2}+(\Vc{\flux}^{-})^{2}}{\Vc{\flux}^{+}+\Vc{\flux}^{-}}$ &  $\Vc{\flux}\circ \tanh^{-1}\left(\frac{\Vc{\flux}}{\Vc{\omega}}\right)$ 
& $\frac{\Vc{\omega}\circ \Vc{\flux}}{\Vc{\omega}^{2}-\Vc{\flux}^{2}}+\tanh^{-1}\left(\frac{\Vc{\flux}}{\Vc{\omega}}\right)$& \mbox{NA}& \mbox{NA}\\
\hline
\end{tabular}
\end{table}

\subsection{Explicit form of value function and HJB equation}
By employing the representation $\EPR_{\Vc{\omega}}(\Vc{\flux})\defeq \langle \Vc{\flux},   \Vc{\psi}_{m}^{-1}\left(\Vc{\flux}/\Vc{\omega}\right)\rangle$ in \eqnref{eq:EPR_j}, \eqnref{eq:opt_k} can be explicitly transformed as
\begin{align}
\Vc{\fluxop}^{\pm}_{t}(\Vc{x};\Vc{\beta}) 
&= \arg\max_{\{\Vc{\flux}^{\pm}\}\in \Real_{\ge 0}^{2 |\setR|}}\left[ \left\langle \left(\stoiMatrix^{\Transpose}\frac{\partial \valueF_{t}(\Vc{x};\Vc{\beta})}{\partial \Vc{x}}+\beta_{2}\Vc{F}(t,\Vc{x})\right),\Vc{\flux}\right\rangle 
    - \EPR_{\Vc{\omega}}(\Vc{\flux})\right]
\end{align}
where we replace the control variable from $\{\Vc{\kcoef}^{\pm}\}$ to $\{\Vc{\flux}^{\pm}\}$ because we can arbitrary modulate $\Vc{\flux}^{\pm}$ by changing $\{\Vc{\kcoef}^{\pm}\}$ for $\Vc{x}\in\Real_{>0}^{|\setX|}$.
By further transforming the control variable from $\{\Vc{\flux}^{\pm}\}$ to $\Vc{\omega}$ and $\Vc{\flux}$, we formally obtain
\begin{align}
(\Vc{\omega}^{\dagger}_{t}(\Vc{x};\Vc{\beta}), \Vc{\flux}^{\dagger}_{t}(\Vc{x};\Vc{\beta}))&= \arg\max_{\Vc{\omega}\in \Real_{\ge 0}^{|\setR|},\Vc{\flux}\in \Real^{|\setR|}}\left[ \left\langle \left(\stoiMatrix^{\Transpose}\frac{\partial \valueF_{t}(\Vc{x};\Vc{\beta})}{\partial \Vc{x}}+\beta_{2}\Vc{F}(t,\Vc{x})\right),\Vc{\flux}\right\rangle 
    - \EPR_{\Vc{\omega}}(\Vc{\flux})\right]. \label{eq:opt_k_2}
\end{align}

Partial optimization of \eqnref{eq:opt_k_2} with respect to $\Vc{\omega}$ for a fixed $\Vc{\flux}$ is equivalent to $\arg \min_{\Vc{\omega}}\EPR_{\Vc{\omega}}(\Vc{\flux})=\langle \Vc{\flux},\Vc{\psi}^{-1}_{\gmean}(\Vc{\flux}/\Vc{\omega})\rangle$. 
Since $\psi^{-1}_{\gmean}(u)$ is odd and monotonous, $\EPR_{\Vc{\omega}}(\Vc{\flux}) \to 0$ as $\Vc{\omega}\to \infty$. Thus, without additional constraints over possible values of $\Vc{\omega}$, the solution becomes singular\cite{dechant2022J.Phys.A:Math.Theor.}.
A representative constraint is directly fixing $\Vc{\omega}$ to a certain value or function $\Vc{\omega}(\Vc{x})$\cite{muratore-ginanneschi2013JStatPhys,yoshimura2023Phys.Rev.Res.a,remlein2021Phys.Rev.E}.
In the main body, we mainly focus on the case where $\Vc{\omega}^{\dagger}_{t}$ is fixed for each $\Vc{x}$ and $t$ as $\Vc{\omega}_{t}(\Vc{x})$. 
Another constraint being investigated before is the total activity budget constraint; $\langle\Vc{1},\Vc{\omega}\rangle\le \Omega$ for  each time point or $\frac{1}{T}\int_{0}^{T}\langle\Vc{1},\Vc{\omega}_{t}\rangle\le \Omega$ over the whole trajectory\cite{dechant2022J.Phys.A:Math.Theor.,vanvu2023Phys.Rev.Xa,nagayama2025Phys.Rev.Res.}, which is treated in Appendix.

Given a fixed $\Vc{\omega}_{t}(\Vc{x})$, we then solve the optimization with respect to $\Vc{\flux}$, which has the (linear)-(convex) Legendre form. 
Thus, the optimal flux $\Vc{\flux}^{\dagger}$ is obtained as
\begin{align}
\Vc{\flux}^{\dagger}_{t}(\Vc{x})&=\arg\max_{\Vc{\flux}}\left[\left\langle\left( \stoiMatrix^{\Transpose} \frac{\partial \valueF_{t}(\Vc{x})}{\partial \Vc{x}}+\beta_{2}\Vc{F}(t,\Vc{x})\right),\Vc{\flux}\right\rangle
    -\EPR_{\Vc{\omega}(\Vc{x})}(\Vc{\flux})\right]\\
    &=\partial_{\Vc{\tff}} \EPR^{*}_{\Vc{\omega}_{t}(\Vc{x})}\left[\stoiMatrix^{\Transpose} \frac{\partial\valueF_{t}(\Vc{x})}{\partial \Vc{x}}+\beta_{2}\Vc{F}(t,\Vc{x})\right]=\partial_{\Vc{\tff}} \EPR^{*}_{\Vc{\omega}_{t}(\Vc{x})}\left[\Vc{\tff}_{t}^{\dagger}(\Vc{x})\right]\label{eq:opt_flux_det},
\end{align}
where the optimal pseudo-force is defined as
\begin{align}
\Vc{\tff}^{\dagger}_{t}(\Vc{x})&\defeq\stoiMatrix^{\Transpose} \frac{\partial\valueF_{t}(\Vc{x})}{\partial \Vc{x}}+\beta_{2}\Vc{F}(t,\Vc{x}).\label{eq:opt_p_force_det} 
\end{align}
From \eqnref{eq:opt_flux_det}, we can see that the optimally controlled dynamics have the form (Fig. \ref{fig:CRN_control} (b)): 
\begin{align}
\frac{\dd \Vc{x}}{\dt}=\stoiMatrix\Vc{\flux}^{\dagger}_{t}(\Vc{x})=\stoiMatrix \partial_{\Vc{\tff}} \EPR^{*}_{\Vc{\omega}_{t}(\Vc{x})}\left[\stoiMatrix^{\Transpose} \frac{\partial\valueF_{t}(\Vc{x})}{\partial \Vc{x}}+\beta_{2}\Vc{F}(t,\Vc{x})\right] \label{eq:opt_dyn_det}
\end{align}
Note that the optimal thermodynamic force  $\Vc{\tf}^{\dagger}_{t}$ is
\begin{align}
\Vc{\tf}^{\dagger}_{t}(\Vc{x})&= \psi^{-1}_{\gmean}\left(\frac{\Vc{\flux}^{\dagger}_{t}(\Vc{x})}{\Vc{\omega}_{t}(\Vc{x})}\right)
= \psi^{-1}_{\gmean}\left(\frac{1}{\Vc{\omega}_{t}(\Vc{x})}\circ \partial_{\Vc{\tff}} \EPR^{*}_{\Vc{\omega}_{t}(\Vc{x})}\left[\stoiMatrix^{\Transpose} \frac{\partial\valueF_{t}(\Vc{x})}{\partial \Vc{x}}+\beta_{2}\Vc{F}(t,\Vc{x})\right]\right)\label{eq:opt_force_det}
\end{align}
and the optimal kinetic parameters are obtained as 
\begin{align}
\Vc{\kcoefop}^{+}_{t}(\Vc{x})&=\Vc{\omega}_{t}\circ  \frac{\psi_{\gmean}(\Vc{\tf}^{\dagger}_{t}(\Vc{x}))}{1-e^{-\Vc{\tf}^{\dagger}_{t}(\Vc{x})}}\circ \frac{1}{\Vc{h}^{+}(\Vc{x})}, 
& \Vc{\kcoefop}^{-}_{t}(\Vc{x})&=\Vc{\omega}_{t}\circ \frac{\psi_{\gmean}(\Vc{\tf}^{\dagger}_{t}(\Vc{x}))}{e^{\Vc{\tf}^{\dagger}_{t}(\Vc{x})}-1}\circ \frac{1}{\Vc{h}^{-}(\Vc{x})}.
\end{align}
To gain an explict form of the HJB equation (\eqnref{eq:HJB_orig_det_after_opt}), we introduce  a Hamiltonian 
\begin{align}
\Hamiltonian_{\Vc{\omega}}[t,\Vc{x},\Vc{\tff},\Vc{\flux}]\defeq \beta_{1}U(t,\Vc{x})+\left[\left\langle\Vc{\flux}, \Vc{\eta}\right\rangle
+\beta_{2}\langle\Vc{F}(t,\Vc{x}),\Vc{\flux}\rangle -\EPR_{\Vc{\omega}}(\Vc{\flux})\right],
\end{align}
Then, the HJB equation after transformation is 
\begin{align}
 -\frac{\partial  \valueF_{t}(\Vc{x})}{\partial t}&= \max_{\Vc{\flux}\in \Real^{R}}\Hamiltonian_{\Vc{\omega}_{t}(\Vc{x})}\left[t, \Vc{x},\stoiMatrix^{\Transpose}\frac{\partial \valueF_{t}(\Vc{x})}{\partial \Vc{x}},\Vc{\flux}\right]
 =\Hamiltonian_{\Vc{\omega}_{t}(\Vc{x})}\left[t,\Vc{x},\stoiMatrix^{\Transpose}\frac{\partial \valueF_{t}(\Vc{x})}{\partial \Vc{x}},\Vc{\flux}^{\dagger}_{t}(\Vc{x})\right]\notag\\
 &=\beta_{1}U(t,\Vc{x}) + \EPR^{*}_{\Vc{\omega}_{t}(\Vc{x})}\left[\stoiMatrix^{\Transpose} \frac{\partial\valueF_{t}(\Vc{x})}{\partial \Vc{x}}+\beta_{2}\Vc{F}(t,\Vc{x})\right]=\Hamiltonian^{\dagger}_{\Vc{\omega}_{t}(\Vc{x})}\left[t,\Vc{x},\stoiMatrix^{\Transpose}\frac{\partial \valueF_{t}(\Vc{x})}{\partial \Vc{x}}\right], \label{eq:opt_HJB}
\end{align}
where the optimal Hamiltonian $\Hamiltonian^{\dagger}$ is defined as
\begin{align}
\Hamiltonian^{\dagger}_{\Vc{\omega}}\left[t,\Vc{x},\Vc{\tff}\right]\defeq\beta_{1}U(t,\Vc{x}) + \EPR^{*}_{\Vc{\omega}}\left[\Vc{\tff}+\beta_{2}\Vc{F}(t,\Vc{x})\right].\label{eq:opt_Hamiltonian}
\end{align}
The corresponding optimal Lagrangian is
\begin{align}
\Lagrangian^{\dagger}_{\Vc{\omega}}[t,\Vc{x},\Vc{\flux}]&\defeq \max_{\Vc{\tff}}\left[\langle \Vc{\flux},\Vc{\tff} \rangle-\Hamiltonian^{\dagger}_{\Vc{\omega}}[t,\Vc{x},\Vc{\tff}]\right]=-\beta_{1}U(t,\Vc{x}) -\beta_{2}\langle\Vc{F}(t,\Vc{x}),\Vc{\flux}\rangle + \EPR_{\Vc{\omega}}[\Vc{\flux}]
\end{align}
The value function along the optimal dynamics (\eqnref{eq:opt_dyn_det}) satisfies
\begin{align}
\frac{\dd \valueF_{t}(t,\Vc{x}_{t})}{\dt}&=\frac{\partial \valueF_{t}(t,\Vc{x}_{t})}{\partial t}+\left\langle \frac{\partial \valueF_{t}(t,\Vc{x}_{t})}{\partial \Vc{x}}, \frac{\dd \Vc{x}}{\dt}\right\rangle\\
&=-\beta_{1}U(t,\Vc{x}) - \EPR^{*}_{\Vc{\omega}_{t}(\Vc{x})}[\Vc{\tff}^{\dagger}_{t}(\Vc{x})]+\left\langle \frac{\partial \valueF_{t}(t,\Vc{x}_{t})}{\partial \Vc{x}},\stoiMatrix \Vc{\flux}^{\dagger}_{t}(\Vc{x})\right\rangle\\
&=-\beta_{1}U(t,\Vc{x}) - \EPR^{*}_{\Vc{\omega}_{t}(\Vc{x})}[\Vc{\tff}^{\dagger}_{t}(\Vc{x})]+\left\langle  \Vc{\tff}^{\dagger}_{t}(\Vc{x})-\beta_{2}\Vc{F}(t,\Vc{x}),\Vc{\flux}^{\dagger}_{t}(\Vc{x})\right\rangle\\
&=-\beta_{1}U(t,\Vc{x})-\beta_{2}\left\langle \Vc{\flux}^{\dagger}_{t}(\Vc{x}), \Vc{F}(t,\Vc{x})\right\rangle+\EPR_{\Vc{\omega}_{t}(\Vc{x})}[\Vc{\flux}^{\dagger}_{t}(\Vc{x})]\\
&=\Lagrangian^{\dagger}_{\Vc{\omega}_{t}(\Vc{x})}[t,\Vc{x},\Vc{\flux}^{\dagger}_{t}(\Vc{x})].
\end{align}
Thus, 
\begin{align}
\valueF_{t}(\Vc{x}_{t})=\beta_{1}\hat{U}_{T}(\Vc{x}_{T})-\int_{t}^{T}\Lagrangian^{\dagger}_{\Vc{\omega}_{t'}(\Vc{x})}[t',\Vc{x}_{t'},\Vc{\flux}^{\dagger}(\Vc{x}_{t'})]\dt'
\end{align}
This is consistent with the original definition of the value function by \eqnref{eq:v_func_path}\footnote{While the sign of the derivative of value function looks opposite, this equation is correct because the time integral of the value function is not $[0,t]$ but $[t,T]$, which induces the flip of the sign.}.

\subsection{Pontryagin's maximum principle and Hamiltonian dynamics}
Although Bellman's principle provides the global perspective on the optimal control problem and the induced dynamics, solving the HJB equation, an infinite-dimensional equation in general, is often computationally prohibitive both analytically and numerically even for low-dimensional CRNs.
Pontryagin's maximum principle offers an alternative way to solve the same control problem via Hamiltonian dynamics, yet locally for each trajectory. 
Specifically, the optimal control solution (\eqnref{eq:opt_dyn_det})  can be casted to the solution of Hamiltonian dynamics. 

By assuming that the optimal value function is smooth, we introduce $\Vc{p}$ as 
\begin{align}
\Vc{p}_{t}\defeq \left.\frac{\partial \valueF_{t}(\Vc{x})}{\partial \Vc{x}}\right|_{\Vc{x}=\Vc{x}^{\dagger}_{t}}
\end{align}
where $\Vc{x}^{\dagger}_{t}$ is the solution of \eqnref{eq:opt_dyn_det}.
Then, the dynamics of $(\Vc{x}_{t},\Vc{p}_{t})$ is determined by the Pontryagin's Hamiltonian dynamics:
\begin{align}
\frac{\dd \Vc{x}}{\dt}&=\quad\frac{\partial \Hamiltonian^{\dagger}_{\Vc{\omega}_{t}(\Vc{x})}[t,\Vc{x},\stoiMatrix^{\Transpose}\Vc{p}]}{\partial \Vc{p}}=\stoiMatrix \partial_{\Vc{\tff}} \EPR^{*}_{\Vc{\omega}_{t}(\Vc{x})}\left[\stoiMatrix^{\Transpose} \Vc{p}+\beta_{2}\Vc{F}(t,\Vc{x})\right]\label{eq:pont_x}\\
\frac{\dd \Vc{p}}{\dt}&=-\frac{\partial \Hamiltonian^{\dagger}_{\Vc{\omega}_{t}(\Vc{x})}[t,\Vc{x},\stoiMatrix^{\Transpose}\Vc{p}]}{\partial \Vc{x}_{t}}= -\beta_{1}\frac{\partial U(t,\Vc{x})}{\partial \Vc{x}} - \frac{\partial}{\partial \Vc{x}}\left\{\EPR^{*}_{\Vc{\omega}_{t}(\Vc{x})}\left[\stoiMatrix^{\Transpose}\Vc{p}+\beta_{2}\Vc{F}(t,\Vc{x})\right]\right\}\label{eq:pont_p}
\end{align}
This finite-dimensional equation solved with appropriate boundary conditions provides  candidates of the optimally controlled trajectory\footnote{The trajectories generated by this equation satisfy only the stationarity condition of the original optimal control problem, which is a necessary but not sufficient optimality condition. Thus, it could have multiple solutions whose optimality should be verified afterward.}.

\subsection{Thermodynamic representation of Pontryagin\textquoteright{}s equation}\label{sec:ThermoRepPont}
Although Hamiltonian dynamics via Pontryagin\textquoteright{}s principle, \eqnref{eq:pont_x}, and \eqnref{eq:pont_p}, is the standard form for solving general optimal control, most of the previous work on thermodynamic transport and control adopted the thermodynamic force $\Vc{\tf}$ rather than $\Vc{\tff}$ as the control variable. 
In addition, the convex conjugate EPR $\EPR^{*}_{\Vc{\omega}}(\Vc{\tff})$ embedded in the equations does not have an analytic representation as a function of $\Vc{\tff}$ for most of the generalized means except the logarithmic one\footnote{This property poses no drawback for numerical computation, given that convex conjugates can be computed efficiently and stably.}.
We resolve these problems by showing that 
$\EPR^{*}_{\Vc{\omega}}$ can be represented analytically as a function of $\Vc{\tf}$, which establishes a thermodynamic representation of Pontryagin\textquoteright{}s equation.
From \eqnref{eq:EPR_j}, we obtain
\begin{align}
\Vc{\eta}=\partial_{\Vc{\flux}} \EPR_{\Vc{\omega}}\left(\Vc{\flux}\right)=\Vc{\psi}^{-1}_{\gmean}\left(\frac{\Vc{\flux}}{\Vc{\omega}}\right)+\frac{\Vc{\flux}}{\Vc{\omega}}\circ \frac{1}{\Vc{\psi}'_{\gmean}\left[\Vc{\psi}^{-1}_{\gmean}\left(\frac{\Vc{\flux}}{\Vc{\omega}}\right)\right]}=\Vc{\tf}+\frac{\Vc{\psi}_{\gmean}(\Vc{\tf})}{\Vc{\psi}'_{\gmean}(\Vc{\tf})} \label{eq:opt_eta_tf}
\end{align}
where $\psi'_{\gmean}(f)\defeq\dd \psi_{\gmean}(f)/\dd f$ (Fig. \ref{fig:CRN_control} (c))\footnote{$\Vc{\tf}+\frac{\Vc{\psi}_{\gmean}(\Vc{\tf})}{\Vc{\psi}'_{\gmean}(\Vc{\tf})}$ is a strictly monotone function, which is derived by the relation among $\Vc{\eta}$, $\Vc{\flux}$ and $\Vc{\tf}$ given $\Vc{\omega}$. This can be derived directly from the second condition in \eqnref{eq:phys_cond}.}. Thus, $\Vc{\eta}-\Vc{\tf}=\frac{\Vc{\psi}_{\gmean}(\Vc{\tf})}{\Vc{\psi}'_{\gmean}(\Vc{\tf})}$ holds.
This relation is also rephrased by using the dissipation function as
\begin{align}
\Vc{\eta}-\Vc{\tf}&=\frac{\partial \EPR(\Vc{\flux})}{\partial\Vc{\flux}}-\Vc{\tf}=\frac{\partial (\dissip_{\Vc{\omega}}(\Vc{\flux})+\dissip^{*}_{\Vc{\omega}}(\Vc{\tf}(\Vc{\flux}))}{\partial\Vc{\flux}}-\Vc{\tf}\\
&=\frac{\partial\dissip^{*}_{\Vc{\omega}}(\Vc{\tf}(\Vc{\flux}))}{\partial\Vc{\flux}}=\frac{\partial \Vc{\tf}(\Vc{\flux})}{\partial \Vc{\flux}}\frac{\partial\dissip^{*}_{\Vc{\omega}}(\Vc{\tf})}{\partial \Vc{\tf}}
=\frac{\partial^{2}\dissip_{\Vc{\omega}}(\Vc{\flux})}{\partial \Vc{\flux}^{2}}\Vc{\flux},
\end{align}
where $\frac{\partial^{2}\dissip_{\Vc{\omega}}(\Vc{\flux})}{\partial \Vc{\flux}^{2}}$ is the Hessian of $\dissip_{\Vc{\omega}}(\Vc{\flux})$.
Then, we have 
\begin{align}
\EPR_{\Vc{\omega}}^{*}(\Vc{\tf})&\defeq\EPR_{\Vc{\omega}}^{*}(\Vc{\eta})=\langle\Vc{\eta},\Vc{\flux} \rangle -\EPR_{\Vc{\omega}}(\Vc{\flux})=\langle\Vc{\eta},\Vc{\flux} \rangle -\langle\Vc{\tf},\Vc{\flux} \rangle=\langle\Vc{\eta}-\Vc{\tf},\Vc{\flux} \rangle\\
&=\left\langle \frac{\Vc{\psi}_{\gmean}(\Vc{\tf})}{\Vc{\psi}'_{\gmean}(\Vc{\tf})}, \Vc{\flux}\right\rangle=\left\langle \frac{\Vc{\psi}_{\gmean}(\Vc{\tf})}{\Vc{\psi}'_{\gmean}(\Vc{\tf})}, \Vc{\omega}\circ \Vc{\psi}_{\gmean}(\Vc{\tf})\right\rangle= \left\langle \Vc{\omega},\frac{\Vc{\psi}_{\gmean}(\Vc{\tf})^{2}}{\Vc{\psi}'_{\gmean}(\Vc{\tf})}\right\rangle \label{eq:EPR_conjugate_f}
\end{align}
where we used \eqnref{eq:opt_eta_tf} from the first to the second lines. Note that, as a function of $\Vc{\flux}$, $\EPR_{\Vc{\omega}}^{*}(\Vc{\eta})$ becomes
\begin{align}
\EPR_{\Vc{\omega}}^{*}(\Vc{\flux})&\defeq\EPR_{\Vc{\omega}}^{*}(\Vc{\eta})=\langle\Vc{\eta}-\Vc{\tf},\Vc{\flux} \rangle=\left\langle \Vc{\flux},\frac{\partial^{2}\dissip_{\Vc{\omega}}(\Vc{\flux})}{\partial \Vc{\flux}^{2}}\Vc{\flux} \right\rangle= \Pi_{\Vc{\omega}}(\Vc{\flux}),
\end{align}
where $\Pi_{\Vc{\omega}}(\Vc{\flux})\defeq \left\langle \Vc{\flux},\frac{\partial^{2}\dissip_{\Vc{\omega}}(\Vc{\flux})}{\partial \Vc{\flux}^{2}}\Vc{\flux} \right\rangle$ is a generalization of the pseudo-entropy production rate (pEPR) via the generalized means\cite{yoshimura2021Phys.Rev.Lett.b,shiraishi2021JStatPhysa,kobayashi2022Phys.Rev.Research}. 
Although pEPR has been recognized as a proxy for EPR when extending the thermodynamic uncertainty relation to MJPs\cite{shiraishi2021JStatPhysa,dechant2022J.Phys.A:Math.Theor.,shiraishi2024J.Stat.Mech.}, our result clarifies that it has a fundamental meaning as the dual of EPR\footnote{The EPR and the pEPR (the dual of EPR) coincide if the EPR is quadratic with respect to $\Vc{\flux}$. This happens in the Langevin processes and MPJs under the logarithmic mean.}. 

Using \eqnref{eq:opt_p_force_det} and \eqnref{eq:opt_eta_tf}, the value function and the optimal force are associated by the following equation:
\begin{align}
\stoiMatrix^{\Transpose} \frac{\partial\valueF_{t}(\Vc{x})}{\partial \Vc{x}}+\beta_{2}\Vc{F}(t,\Vc{x})=\Vc{\tff}^{\dagger}_{t}(\Vc{x})=\Vc{\tf}^{\dagger}_{t}(\Vc{x})+\frac{\Vc{\psi}_{\gmean}(\Vc{\tf}^{\dagger}_{t}(\Vc{x}))}{\Vc{\psi}'_{\gmean}(\Vc{\tf}^{\dagger}_{t}(\Vc{x}))}.\label{eq:opt_forces}
\end{align}
The HJB equation (\eqnref{eq:opt_HJB}) is transformed into 
\begin{align}
-\frac{\partial  \valueF_{t}(\Vc{x})}{\partial t}&=\beta_{1}U(t,\Vc{x}) + \EPR^{*}_{\Vc{\omega}_{t}(\Vc{x})}\left[\Vc{\tff}^{\dagger}_{t}(\Vc{x})\right]=\beta_{1}U(t,\Vc{x}) + \left\langle \Vc{\omega}_{t}(\Vc{x}),\frac{\Vc{\psi}_{\gmean}(\Vc{\tf}^{\dagger}_{t}(\Vc{x}))^{2}}{\Vc{\psi}'_{\gmean}(\Vc{\tf}^{\dagger}_{t}(\Vc{x}))}\right\rangle 
\end{align}
where we used \eqnref{eq:EPR_conjugate_f}.
For Pontryagin\textquoteright{}s equation, $\Vc{p}$ and $\Vc{\tf}$ are associated as
\begin{align}
\stoiMatrix^{\Transpose} \Vc{p}+\beta_{2}\Vc{F}(t,\Vc{x})=\Vc{\eta}=\Vc{\tf}+\frac{\Vc{\psi}_{\gmean}(\Vc{\tf})}{\Vc{\psi}'_{\gmean}(\Vc{\tf})},\label{eq:opt_eta_tf_2}
\end{align}
By using \eqnref{eq:j_f} and \eqnref{eq:EPR_conjugate_f}, \eqnref{eq:pont_x} and \eqnref{eq:pont_p} become
\begin{align}
\frac{\dd \Vc{x}}{\dt}&=\stoiMatrix[\Vc{\omega}_{t}(\Vc{x})\circ \Vc{\psi}_{\gmean}(\Vc{\tf})], 
\label{eq:pont_x_f}\\
\frac{\dd \Vc{p}}{\dt}&=-\beta_{1}\frac{\partial U(t,\Vc{x})}{\partial \Vc{x}} - \beta_{2}\left(\frac{\partial \Vc{F}(t,\Vc{x})}{\partial \Vc{x}}\right)^{\Transpose}\Vc{\omega}_{t}(\Vc{x})\circ \Vc{\psi}_{\gmean}(\Vc{\tf}) -\left(\frac{\partial \Vc{\omega}_{t}(\Vc{x})}{\partial \Vc{x}}\right)^{\Transpose}\frac{\Vc{\psi}_{\gmean}(\Vc{\tf})^{2}}{\Vc{\psi}'_{\gmean}(\Vc{\tf})}\label{eq:pont_p_f}
\end{align}
By differentiating \eqnref{eq:opt_eta_tf_2} along $\Vc{x}_{t}^{\dagger}$ with respect to $t$, we have
\begin{align}
\diag\left[\Vc{2}-\frac{\psi_{\gmean}(\Vc{\tf})\circ\psi''_{\gmean}(\Vc{\tf})}{(\psi'_{\gmean}(\Vc{\tf}))^{2}}\right]\frac{\dd\Vc{\tf}}{\dt}=\stoiMatrix^{\Transpose}\frac{\dd \Vc{p}}{\dt}+\beta_{2}\left[\frac{\partial\Vc{F}(t,\Vc{x})}{\partial t}+\left(\frac{\partial \Vc{F}(t,\Vc{x})}{\partial \Vc{x}}\right)\frac{\dd \Vc{x}}{\dt}\right]
\end{align}
By inserting \eqnref{eq:pont_x_f} and \eqnref{eq:pont_p_f}, we have
\begin{align}
\frac{\dd\Vc{\tf}}{\dt}&=\diag\left[\Vc{2}-\frac{\psi_{\gmean}(\Vc{\tf})\circ\psi''_{\gmean}(\Vc{\tf})}{(\psi'_{\gmean}(\Vc{\tf}))^{2}}\right]^{-1}\left\{-\beta_{1}\stoiMatrix^{\Transpose}\frac{\partial U(t,\Vc{x})}{\partial \Vc{x}}-\stoiMatrix^{\Transpose}\left(\frac{\partial \Vc{\omega}_{t}(\Vc{x})}{\partial \Vc{x}}\right)^{\Transpose}\frac{\Vc{\psi}_{\gmean}(\Vc{\tf})^{2}}{\Vc{\psi}'_{\gmean}(\Vc{\tf})}\right.\notag\\
&+\left.\beta_{2}\left[\frac{\partial\Vc{F}(t,\Vc{x})}{\partial t}+\left\{\left(\frac{\partial \Vc{F}(t,\Vc{x})}{\partial \Vc{x}}\stoiMatrix\right) -\left(\frac{\partial \Vc{F}(t,\Vc{x})}{\partial \Vc{x}}\stoiMatrix\right)^{\Transpose} \right\}\Vc{\omega}_{t}(\Vc{x})\circ \Vc{\psi}_{\gmean}(\Vc{\tf})\right]\right\}\label{eq:pont_f_f}
\end{align}
We also have an algebraic constraint  
\begin{align}
\Vc{\tf}+\frac{\Vc{\psi}_{\gmean}(\Vc{\tf})}{\Vc{\psi}'_{\gmean}(\Vc{\tf})}-\beta_{2}\Vc{F}(t,\Vc{x})=\stoiMatrix^{\Transpose} \Vc{p}\in \Img \stoiMatrix^{\Transpose},\label{eq:pont_f_const}
\end{align}
which originates from \eqnref{eq:opt_eta_tf} and accounts for the difference of dimensions of $\Vc{p}$ and $\Vc{\tf}$.
The set of equations, \eqnref{eq:pont_x_f}, \eqnref{eq:pont_f_f} and \eqnref{eq:pont_f_const} establish  the thermodynamic version of Pontryagin\textquoteright{}s Hamiltonian dynamics.

\section{Analytic solutions and its relationship with thermodynamic transport problems}\label{sec:analytic}
In this section, we demonstrate that the transport problem with the constant flux utility is a theoretically sound extension that unifies the usual thermodynamic transport problem and the flux optimization problem. 
Specifically, we consider the setup such that $U(t,\Vc{x})=0$, and $\beta_{2}\Vc{F}(t,\Vc{x})=\Vc{F}$, and the boundary conditions are $\Vc{x}_{0}=\Vc{x}_{init}$ and $\Vc{x}_{T}=\Vc{x}_{end}$, satisfying  $\Vc{x}_{init}-\Vc{x}_{end}\in \Img \stoiMatrix$. Then, the thermodynamic Pontryagin equation for $\Vc{\tf}$ is reduced to  
\begin{align}
\frac{\dd\Vc{\tf}}{\dt}&=-\diag\left[\Vc{2}-\frac{\psi_{\gmean}(\Vc{\tf})\psi''_{\gmean}(\Vc{\tf})}{(\psi'_{\gmean}(\Vc{\tf}))^{2}}\right]^{-1}\left\{\stoiMatrix^{\Transpose}\left(\frac{\partial \Vc{\omega}_{t}(\Vc{x})}{\partial \Vc{x}}\right)^{\Transpose}\frac{\Vc{\psi}_{\gmean}(\Vc{\tf})^{2}}{\Vc{\psi}'_{\gmean}(\Vc{\tf})}\right\}.
\label{eq:TPE_transport}
\end{align}
While solving this equation together with \eqnref{eq:pont_x_f} and \eqnref{eq:pont_f_const} analytically is still impossible in general, it becomes tractable if the right-hand side of \eqnref{eq:TPE_transport} becomes $0$ or effectively $0$. 
Most previous works for thermodynamic transport exploit such situations\cite{dechant2022J.Phys.A:Math.Theor.,yoshimura2023Phys.Rev.Res.a,vanvu2023Phys.Rev.Xa,nagayama2025Phys.Rev.Res.}.

\subsection{Optimal thermodynamic transports with flux utility under constant activity and with total activity constraints}
If the activity $\Vc{\omega}_{t}(\Vc{x})$ is constant $\Vc{\omega}$, the thermodynamic Pontryagin's equation substantially simplifies to
\begin{align}
\frac{\dd \Vc{x}}{\dt}&=\stoiMatrix\Vc{\omega}\circ \Vc{\psi}_{\gmean}(\Vc{\tf}), & 
\frac{\dd\Vc{\tf}}{\dt}&=0, & \Vc{\tf}&+\frac{\Vc{\psi}_{\gmean}(\Vc{\tf})}{\Vc{\psi}'_{\gmean}(\Vc{\tf})}-\Vc{F}\in  \Img \stoiMatrix^{\Transpose}
\end{align}
Then, the optimal trajectories are formally $\Vc{\tf}^{\dagger}(t)=\Vc{\tf}^{\dagger}=\const$ and $\Vc{x}^{\dagger}(t)=\Vc{x}_{init}+t\Vc{v}^{\dagger}$ where $\Vc{v}^{\dagger}\defeq (\Vc{x}_{end}-\Vc{x}_{init})/T$.
If $\Vc{v}^{\dagger} \in \{\stoiMatrix\Vc{\flux}|\flux_{r}/\omega_{r}\in D^{u}\,\mbox{for all } r\in \setR \}$ holds, the optimal constant force $\Vc{\tf}^{\dagger}$ is obtained by solving  $\stoiMatrix\Vc{\omega}\circ \Vc{\psi}_{\gmean}(\Vc{\tf}^{\dagger})=\stoiMatrix\Vc{\flux}^{\dagger}= \Vc{v}^{\dagger}$ and $\Vc{\tff}^{\dagger}-\Vc{F}=\Vc{\tf}^{\dagger}+\frac{\Vc{\psi}_{\gmean}(\Vc{\tf}^{\dagger})}{\Vc{\psi}'_{\gmean}(\Vc{\tf}^{\dagger})}-\Vc{F}\in  \Img \stoiMatrix^{\Transpose}$. These equations are equivalent to $\Vc{\flux}^{\dagger}\in \Ker \stoiMatrix + \Vc{\flux}'$ and $\Vc{\tff}^{\dagger}=\partial \EPR_{\Vc{\omega}}(\Vc{\flux}^{\dagger})\in \Img \stoiMatrix^{\Transpose}+\Vc{F}$ where $\Vc{\flux}'$ is a particular solution of $\stoiMatrix\Vc{\flux}=\Vc{v}^{\dagger}$.
Since $\Img \stoiMatrix^{\Transpose}$ and $\Ker \stoiMatrix$ are orthogonal complement, the optimal $\Vc{\flux}^{\dagger}$ is uniquely determined as $\Vc{\flux}^{\dagger}=(\Ker \stoiMatrix + \Vc{\flux}')\cap \partial \EPR^{*}_{\Vc{\omega}}[\Img\stoiMatrix^{\Transpose}+\Vc{F}]$ by the information-geometric orthogonarity\cite{kobayashi2024Info.Geo.}.
Then $\Vc{\tf}^{\dagger}$ and $\Vc{\tff}^{\dagger}$ are uniquely determined from $\Vc{\flux}^{\dagger}$.
For this setup, the value function is obtained analytically as
\begin{align}
\valueF_{t}(\Vc{x})=-(T-t)\hat{\Lagrangian}^{\dagger}_{\Vc{\omega}}\left(\frac{\Vc{x}_{end}-\Vc{x}}{T-t} \right)
\end{align}
where $\hat{\Lagrangian}^{\dagger}_{\Vc{\omega}}(\Vc{v})$ is the contracted (optimal) Lagrangian defined as 
\begin{align}
\hat{\Lagrangian}^{\dagger}_{\Vc{\omega}}(\Vc{v})\defeq 
\min_{\Vc{v}=\stoiMatrix\Vc{\flux}}\Lagrangian^{\dagger}_{\Vc{\omega}}[\Vc{\flux}]=\begin{cases}
    \min_{\Vc{v}=\stoiMatrix\Vc{\flux}}\left[\EPR_{\Vc{\omega}}(\Vc{\flux)}-\langle\Vc{F}, \Vc{\flux} \rangle\right]. & \mbox{if $\Vc{v}\in \Img \stoiMatrix$}\\
    + \infty & \mbox{otherwise}
\end{cases}
\end{align}
(see Appendix for derivation). 

Yoshimura \textit{et al.} considered this situation for $\Vc{F}=0$ by using the logarithmic mean and directly constraining $\Vc{\omega}$ to be constant\cite{yoshimura2023Phys.Rev.Res.a}. In such a case, the optimal force becomes conservative because $\Vc{\tff}^{\dagger}=\Vc{\tf}^{\dagger}+\frac{\Vc{\psi}_{\gmean}(\Vc{\tf}^{\dagger})}{\Vc{\psi}'_{\gmean}(\Vc{\tf}^{\dagger})}=2\Vc{\tf}^{\dagger}\in  \Img \stoiMatrix^{\Transpose}$ for the logarithmic mean.
Also, the optimal flux $\Vc{\flux}^{\dagger}=\Vc{\omega}\circ \Vc{\tf}^{\dagger}$ is the dual of $\Vc{\tf}^{\dagger}$ w.r.t. the Riemannian metric $\diag \Vc{\omega}$.

In addition, our formulation includes the flux maximization at the steady state\cite{fleming2012JournalofTheoreticalBiology,gagrani2026J.Stat.Mech.,pinero2024Phys.Rev.Res.,bilancioni2025NatCommun} as a special case 
Specifically, if $\Vc{x}_{init}=\Vc{x}_{end}$, the value function reduces to $\valueF_{0}(\Vc{x}_{init})=-T\hat{\Lagrangian}^{\dagger}_{\Vc{\omega}}\left(\Vc{0} \right)=T \max_{\stoiMatrix\Vc{\flux}=\Vc{0}}\left[\langle\Vc{F}, \Vc{\flux}\rangle-\EPR_{\Vc{\omega}}(\Vc{\flux)} \right]$, which is equivalent to the maximization of the steady flux with the EPR cost.

A similar result is also obtained by minimizing the entropy production under the constraint that the time-average total activity budget is bounded:
\begin{align}
\frac{1}{T}\int_{0}^{T}\dt' \langle \Vc{1},\Vc{\omega}_{t'}(\Vc{x}_{t'})\rangle\le \Omega
\end{align}
The same argument applies as before; the value function becomes
\begin{align}
\valueF_{t}(\Vc{x},m)=-m \hat{\Lagrangian}^{\dagger}\left(\frac{\Vc{x}_{end}-\Vc{x}}{m} \right).
\end{align}
where, for $h(u)=u\psi^{-1}_{\gmean}(u)$,  
\begin{align}
\hat{\Lagrangian}^{\dagger}(\Vc{v})\defeq 
\begin{cases}
    \min_{\Vc{v}=\stoiMatrix\Vc{\xi}}\left[ h\left(\|\Vc{\xi}\|_{1}\right)-\langle\Vc{F}, \Vc{\xi} \rangle\right]. & \mbox{if $\Vc{v}\in \Img \stoiMatrix$}\\
    + \infty & \mbox{otherwise}
\end{cases}
\end{align}
and $m\ge 0$ is the remaining activity budget at $t$ whose initial condition at $t=0$ is $m_{0}=T\Omega$. 
If we reduce the result to the conventional transport problem without flux utility, i.e.,$\Vc{F}=\Vc{0}$, then the value function is further simplified as 
\begin{align}
\valueF_{t}(\Vc{x},m)=-m h\left(\frac{\mathcal{V}_{\stoiMatrix}(\Vc{x}_{end}-\Vc{x})}{m} \right)
\end{align}
where 
\begin{align}
\mathcal{V}_{\stoiMatrix}(\Vc{v})\defeq 
\begin{cases}    \min_{\Vc{\xi}:\stoiMatrix\Vc{\xi}=\Vc{v}}\|\Vc{\xi}\|_{1} & \mbox{if $\Vc{v}\in\Img \stoiMatrix$}\\
+ \infty & \mbox{otherwise.}
\end{cases}
\end{align}
is a norm on $\Img \stoiMatrix$. 
This situation has been investigated for  the arithmetic mean\cite{dechant2022J.Phys.A:Math.Theor.}, the logarithmic mean\cite{vanvu2023Phys.Rev.Xa}, and the general means\cite{nagayama2025Phys.Rev.Res.}.
If this result is applied to MJP where $\stoiMatrix$ is replaced with the graph incidence matrix $\IncMatrix$, $\mathcal{W}(\Vc{x},\Vc{x}')\defeq \mathcal{V}_{\stoiMatrix}(\Vc{x}'-\Vc{x})$ becomes the graph $L^{1}$-Wasserstein distance. 

Since optimizing $\Vc{\omega}$ under this constraint complicates the derivation of the HJB and Pontryagin's equations, we defer the details to Appendix.

\subsection{Optimal thermodynamic transports with \texorpdfstring{$\Vc{x}$}{x}--dependent activity}
If $\Vc{\omega}$ is $\Vc{x}$-dependent, the dynamics of force (\eqnref{eq:TPE_transport}) is no longer constant. 
Muratore-Ginanneschi \textit{et al.}, in their pioneering work, investigated the optimal thermodynamic transport over MJP with $\Vc{x}$-dependent $\Vc{\omega}(\Vc{x})$ where $\Vc{\omega}(\Vc{x})$ is specified by the geometric mean of MA linear kinetics\cite{muratore-ginanneschi2013JStatPhys}.
Remlein and Seifert employed this result to show that non-conservative force is optimal under this choice of activity\cite{remlein2021Phys.Rev.E}.
This choice can be interpreted as defining an activity field based on uncontrolled MA kinetics as a reference. While an analytic expression of the value function and the optimal trajectory cannot be obtained for this $\Vc{x}$-dependent activity in general, we can immediately obtain the main results of \cite{muratore-ginanneschi2013JStatPhys} for the generalized mean via our formulation.
Specifically, the optimal force condition (Eq. 52b in \cite{muratore-ginanneschi2013JStatPhys}), the dynamics of auxiliary field (Eq. 54 in \cite{muratore-ginanneschi2013JStatPhys}), and the dynamics of optimal force (Eq. 56 in in \cite{muratore-ginanneschi2013JStatPhys}) are equivalent to \eqnref{eq:opt_forces}, \eqnref{eq:pont_p_f}, and \eqnref{eq:pont_f_f}, respectively (See Appendix for the detailed comparison). 

In addition, the non-conservativeness of the optimal thermodynamic force for generalized means other than the logarithmic one\cite{remlein2021Phys.Rev.E} is a consequence of \eqnref{eq:opt_forces}, which means that the optimal force $\Vc{\tf}^{\dagger}_{t}(\Vc{x})$ is a nonlinear function of the gradient of the value function $\Vc{\tf}^{\dagger}_{t}(\Vc{x})+\frac{\Vc{\psi}_{\gmean}(\Vc{\tf}^{\dagger}_{t}(\Vc{x}))}{\Vc{\psi}'_{\gmean}(\Vc{\tf}^{\dagger}_{t}(\Vc{x}))}=\stoiMatrix^{\Transpose}\frac{\partial\valueF_{t}(\Vc{x})}{\partial \Vc{x}}$ and thus non-conservative\footnote{See Appendix for more details.}.
Dechant \& Meer proposed to replace the EPR, $\EPR_{\Vc{\omega}(\Vc{x})}(\Vc{\flux})$, with twice the dissipation function $2\Psi_{\Vc{\omega}(\Vc{x})}(\Vc{\flux})$ as the control cost to recover the conservativeness of the optimal force\cite{dechant2026}, the underlying idea of which is close to the variational representation of thermodynamic flux\cite{kobayashi2024Info.Geo.}.
In contrast, our result clarified that the optimal pseudo-force is always the gradient of the value function for $\Vc{F}=\Vc{0}$ as $\Vc{\tff}^{\dagger}_{t}(\Vc{x})=\stoiMatrix^{\Transpose} \frac{\partial\valueF_{t}(\Vc{x})}{\partial \Vc{x}}$, which shares the same dynamical property as the equilibrium dynamics(\eqnref{eq:CRN_var}).
Since the transient dynamics has non-negative EPR even if driven by a conservative force, the generalized gradient flow property is more fundamental for characterizing the optimal thermodynamic transport dynamics.
 (See Appendix for more detailed comparison and discussion).

In contrast to the geometric mean, the logarithmic mean is not compatible with $\Vc{x}$-dependent activity because $\Vc{\omega}(\Vc{x})$ becomes a complicated function even if the simplest linear kinetics is employed.

\section{Conclusion and Discussion}\label{sec:conclusion}
In this work, we have established a thermodynamic optimal control theory for CRNs that aligns with their underlying thermodynamic and geometric structures. By treating reaction rate constants as control variables and incorporating both state and flux utilities alongside the EPR cost, this framework encompasses conventional thermodynamic transport far from equilibrium as well as a broader class of nonequilibrium control tasks.

Our central result clarifies that the optimal dynamics derived from the thermodynamic Hamilton--Jacobi--Bellman (HJB) equation takes the form of the perturbed generalized gradient flow driven by the value function. This characterization stems from the convex representation of the EPR function via generalized means, where its convex conjugate naturally acts as a nonlinear generalization of the friction metric. The resulting dynamics shares the thermodynamic structure of equilibrium and nonequilibrium CRNs and admits an information-geometric characterization. 
Furthermore, we derive a thermodynamic Pontryagin equation by recasting the Pontryagin Hamiltonian system using thermodynamic forces as the control variables. This formulation highlights the intrinsic thermodynamic and geometric structure of the control problem and clarifies its direct connections to optimal transport and related approaches.


These results lay a theoretical foundation for the thermodynamic control of CRNs, with potential applications to biological systems and finite-time thermodynamics.
In addition, they pave the way for addressing the following challenges.

\subsection{From transport to time-cyclic control}
The first important problem for future work is the design of optimal thermodynamic cycles. Although thermodynamic transport between two prescribed state distributions at minimal thermodynamic cost is a basic building block of thermodynamic control, constructing optimal thermodynamic engines requires identifying time-periodic control trajectories that balance dissipation minimization with utility maximization\cite{brandner2015Phys.Rev.X,wareham2026}. Within near-equilibrium thermodynamic geometry, this task can be formulated as a geometric  problem to a finding optimal closed curve on a manifold through an isoperimetric formulation\cite{brandner2020Phys.Rev.Lett.,abiuso2020Phys.Rev.Lett.,frim2022Phys.Rev.Lett.}. Extensions to far-from-equilibrium regimes have also been explored\cite{movillamiangolarra2021Phys.Rev.E,movillamiangolarra2022IEEEControlSyst.Lett.,venturasiches2024Phys.Rev.E,zhong2026Phys.Rev.E}.

Building upon the framework developed here,  a natural next step is to investigate this problem from a geometric perspective, establish conditions for the existence of optimal periodic trajectories, and characterize their thermodynamic properties in more general settings. In parallel, practical numerical schemes for computing optimal cycles should be developed, leveraging the rich toolkit of modern control theory.

\subsection{From deterministic to stochastic control}
A second challenge lies in extending our framework to stochastic optimal control. The conventional  thermodynamic transport and control based on thermodynamic or Wasserstein geometry fall within the class of deterministic optimal control, as they treat control problems for probability distributions or densities governed by deterministic evolution equations, i.e., the Fokker--Planck and master equations. 
Our results for CRNs, which generalize these approaches, belong likewise to this category. Owing to the deterministic nature, an experimental implementation of such control strategies does not require feedback control via observing the systems\cite{oikawa2025NatCommun}.

However, recent advances in single-molecule techniques, such as imaging and manipulation, enable us to access individual stochastic trajectories and directly perturb their dynamics\cite{toyabe2010NaturePhys,ribezzi-crivellari2019Nat.Phys.,saha2022Phys.Rev.Lett.}. Bridging this gap between theory and experiment calls for extending our framework from deterministic to stochastic optimal control. Adapting the established theory of stochastic optimal control\cite{fleming1975,horiguchi2025PRXLife} to thermodynamic settings therefore represents a crucial frontier for future research\cite{reinalter2026}.

\subsection{From nonautonomous to autonomous control}
Third, although external control is the standard setup for studying thermodynamic systems, real heat engines and living systems operate autonomously\cite{serra-garcia2016Phys.Rev.Lett.,jarzynski2025Proc.Natl.Acad.Sci.}. Designing autonomous heat engines and understanding the thermodynamics of living systems therefore require identifying optimal control laws implemented within the systems themselves\cite{yang2021Proc.Natl.Acad.Sci.,thutupalli2024Nat.Phys.}.

Automatic differentiation and learning theory have been employed to obtain such autonomous systems\cite{engel2023Phys.Rev.X,yan2022Phys.Rev.E,wareham2026}.
Among others, reinforcement learning (RL), which builds on optimal control theory, provides methods for formulating and optimizing time-invariant feedback policies\cite{sutton2018a}. The integration of RL with our theory would provide a systematic framework for autonomous thermodynamic control\cite{whitelam2023Phys.Rev.X}.

\subsection{Applications to physics of intelligence}
Finally, the growing interest in the physics of intelligence, spurred in part by advances in artificial intelligence, has stimulated efforts to employ stochastic thermodynamics and related methods to information science\cite{kodama2025,dambrosia2026IEEEJ.Sel.AreasInf.Theory,sagawa2026,ding2026,aslyamov2026a,hasegawa2026}. For example, recent studies have drawn on stochastic thermodynamics to assess the performance of generative diffusion models\cite{ikeda2025Phys.Rev.X}. These models exploit time reversal to learn generative dynamics, while optimal control theory offers a more general formulation by seeking control laws that generate reverse-time trajectories. 
This control-theoretic perspective is applied to state-of-the-art diffusion models and their fine-tuning. Extending the theoretical framework developed here beyond thermodynamics to diverse problems in information science could thus open new avenues for physics of intelligence.

\section{Acknowledgement}
We used OpenAI's ChatGPT for literature review, scientific discussions, and English editing and proofreading. 
All AI-assisted content was independently verified by the authors, who take full responsibility for the manuscript.

We thank Praful Gagrani, Dimitri Loutchko, and Keisuke Sugie for fruitful discussion and comments.
This work was supported by JST CREST (Grant Numbers JPMJCR25Q2), JST PRESTO (Grant No. JPMJPR25KB), and JSPSKAKENHI (Grant Number 25H01365).

%

\end{document}